\documentclass[11pt,a4paper]{article}
\usepackage{jcappub}
\pdfoutput=1
\usepackage[english]{babel}
\usepackage{amsmath,amssymb,amsfonts, bm,bbm,slashed, subdepth}
\usepackage{graphicx}
\usepackage[sort&compress]{natbib}
\usepackage{xcolor}
\usepackage[normalem]{ulem}
\usepackage{hyperref}
\usepackage{cleveref}
\definecolor{red}{rgb}{1.0, 0, 0}
\usepackage{hyperref}
\usepackage{enumerate}
\usepackage{epsfig, subfigure}
\usepackage{setspace}
\usepackage{booktabs, tabularx}
\usepackage{units}
\usepackage{hyperref}

\newcommand{\be}{\begin{equation}}
\newcommand{\ee}{\end{equation}}
\newcommand{\ba}{\begin{array}}
\newcommand{\ea}{\end{array}}
\newcommand{\bea}{\begin{eqnarray}}
\newcommand{\eea}{\end{eqnarray}}
\newcommand{\balg}{\begin{align}}
\newcommand{\ealg}{\end{align}}
\newcommand{\bit}{\begin{itemize}}
\newcommand{\eit}{\end{itemize}}
\newcommand{\trm}[1]{\textrm{#1}}

\usepackage{enumitem}

\newcommand{\Mpc}{\trm{\Mpc}}
\newcommand{\yr}{\trm{\yr}}
\newcommand{\eV}{\trm{\eV}}

\begin{document}
\title{N-body simulations of structure formation in thermal inflation cosmologies}


\author[a]{Matteo Leo,}
\author[b]{Carlton M. Baugh,}
\author[b]{Baojiu Li}
\author[a]{and Silvia Pascoli}


\affiliation[a]{Institute for Particle Physics Phenomenology, Department of Physics, Durham University, South Road, Durham DH1 3LE, U.K.}
\affiliation[b]{Institute for Computational Cosmology, Department of Physics, Durham University, South Road, Durham DH1 3LE, U.K.}

\emailAdd{matteo.leo@durham.ac.uk}
\emailAdd{c.m.baugh@durham.ac.uk} 
\emailAdd{baojiu.li@durham.ac.uk}
\emailAdd{silvia.pascoli@durham.ac.uk}
\hfill{IPPP/18/57}

\abstract{Thermal inflation models (which feature two inflationary stages) can display damped primordial curvature power spectra on small scales which generate damped matter fluctuations.  For a reasonable choice of parameters, thermal inflation models naturally predict a suppression of the matter power spectrum on galactic and sub-galactic scales,   mimicking the effect of warm or interacting dark matter. Matter power spectra in these models are also characterised by an excess of power (with respect to the standard $\Lambda$CDM power spectrum) just below the suppression scale.  By running a suite of N-body simulations we investigate the non-linear growth of structure in models of thermal inflation. We measure the non-linear matter power spectrum and extract halo statistics, such as the halo mass function, and compare these quantities with those predicted in the standard $\Lambda$CDM model and in other models with damped matter  fluctuations. We find that the thermal inflation models considered here produce measurable differences in the matter power spectrum from $\Lambda$CDM at redshifts $z>5$ for wavenumbers $k\in[2,64]\,h\,\mathrm{Mpc}^{-1}$, while the halo mass functions are appreciably different at all redshifts in the halo mass range $M_\mathrm{halo}\in [10^9,10^{12}] \,h^{-1}\,\mathrm{M}_\odot$ resolved by our simulations. The halo mass function at $z=0$ for thermal inflation displays an enhancement of around $\sim 20\%$ with respect to $\Lambda$CDM and a damping at lower halo masses, with the position of the enhancement depending on the value of the free parameter in the model. The enhancement in the halo mass function (with respect to $\Lambda$CDM ) increases with redshift, reaching $\sim 40\%$ at $z=5$. We also study the accuracy of the analytical Press-Schechter approach, using different filters to smooth the density field, to predict halo statistics for thermal inflation. We find that the  predictions with the smooth-$k$ filter we proposed in a separate paper agree with the simulation results over a wider range of halo masses than is the case with other filters commonly used in the literature.}

\maketitle

\section{Introduction}
\label{sec:intro}
The standard cosmological paradigm (standard $\Lambda$CDM hereafter) has proved to be a successful theory that is able to reproduce observations on large scales. This model is characterized by (i) a nearly scale-invariant primordial curvature power spectrum and by (ii) cold and non-interacting dark matter. These two properties of the standard paradigm mean that matter density fluctuations are non-vanishing on all scales.  However, some possible failures have been identified in the standard $\Lambda$CDM model at galactic and sub-galactic scales (e.g. the mismatch between the observed and expected numbers of satellites in the Milky Way; for a review of the small scale problems of the standard paradigm see \cite{Weinberg:2013aya}), although it is not clear if these issues can be resolved within the standard paradigm by considering e.g. baryonic effects \cite{2012MNRAS.421.3464P,2013MNRAS.432.1947M,2014ApJ...786...87B,2012MNRAS.424.2715W}. These failures have renewed interest in alternative scenarios which display less power on small scales than $\Lambda$CDM. 

Damped matter fluctuations can be achieved by relaxing one of the above assumptions characterizing the standard paradigm. We can then divide the models with damped matter fluctuations ({\it damped models} hereafter) into two broad classes: those involving modifications in the primordial power spectrum (e.g. broken scale invariance during inflation, which we dub {\it non-standard inflation} models) \cite{Kamionkowski:1999vp,White:2000sy,Yokoyama:2000tz,Zentner:2002xt,Ashoorioon:2006wc,Kobayashi:2010pz,Nakama:2017ohe,Hong:2015oqa,Starobinsky:1992ts} and those that suppress power at later times through some non-standard DM mechanisms (these models are generally referred to as {\it non-cold dark matter} or nCDM, see e.g. \cite{Murgia:2017lwo}) \cite{Bode:2000gq,Colin:2000dn,Hansen:2001zv,Viel:2005qj,Dodelson:1993je,Dolgov:2000ew,Asaka:2006nq,Enqvist:1990ek,Shi:1998km,Abazajian:2001nj,Kusenko:2006rh,Petraki:2007gq,Merle:2015oja,Konig:2016dzg,Boehm:2004th,Boehm:2014vja,Schewtschenko:2014fca,Spergel:1999mh,Marsh:2015xka,Veltmaat:2016rxo,Veltmaat:2018dfz}. Non-standard inflation models are characterized by a suppression in the primordial curvature power spectrum on small scales (which acts as the seed of all the density perturbations), while the DM sector remains the same as in the standard paradigm (in these models the DM particles are still cold and non-interacting). A suppression in the curvature power spectrum can be achieved e.g. when the first derivative of the inflaton potential (in one-field inflation models) has a discontinuity  \cite{Starobinsky:1992ts, Nakama:2017ohe,Kamionkowski:1999vp} or when a second inflationary stage is introduced (as in models of thermal inflation, see below) \cite{Hong:2015oqa}. nCDM models, on the other hand, introduce non-standard DM mechanisms that modify the shape of the power spectrum during the evolution of the fluctuations in the radiation and matter domination epochs, while the primordial power spectrum is the standard scale-invariant one. The mechanism leading to a suppression of power in nCDM depends on the particular particle production process. Nevertheless, nCDM candidates are often characterized either by a non-negligible thermodynamic velocity dispersion (the so-called {\it warm DM} models or WDM \cite{Bode:2000gq,Colin:2000dn,Hansen:2001zv,Viel:2005qj,Dodelson:1993je,Dolgov:2000ew,Asaka:2006nq,Enqvist:1990ek,Shi:1998km,Abazajian:2001nj,Kusenko:2006rh,Petraki:2007gq,Merle:2015oja,Konig:2016dzg}), interactions (DM interacting with standard model particles such as neutrinos or photons \cite{Boehm:2004th,Boehm:2014vja,Schewtschenko:2014fca} and self-interacting DM \cite{Spergel:1999mh}) or pressure terms from macroscopic wave-like behaviour (e.g. ultra-light axions \cite{Marsh:2015xka,Veltmaat:2016rxo,Veltmaat:2018dfz}).

Historically, thermal inflation was introduced to solve the moduli problem \cite{Lyth:1995hj,Lyth:1995ka}. The moduli are long-lived scalar fields generally present in supersymmetric models. Due to their properties, moduli can dominate the energy density of the Universe for a sufficiently long time to interfere with the epoch of big bang nucleosynthesis (BBN) (this is referred to as the cosmological moduli problem) \cite{Banks:1993en,deCarlos:1993wie}. Thermal inflation solves this problem by introducing a second, low-energy inflationary period that dilutes the moduli density to harmless values. The second inflation period is induced by a new field (the so-called {\it flaton}) trapped at its origin by coupling with the thermal bath \cite{Lyth:1995hj,Lyth:1995ka}. Thermal inflation ends when the temperature is no longer sufficiently high to maintain the flaton at $\phi = 0$, so the field rolls toward its minimum and starts to oscillate, giving rise to a flaton matter dominated period. Finally, the flaton decays, ensuring the standard radiation-domination period before BBN. 

It was recently pointed out that models of thermal inflation can produce interesting effects on the matter density perturbations \cite{Hong:2015oqa}. Indeed, in thermal inflation the standard inflationary stage is followed by additional periods that can modify the nearly scale-invariant curvature power spectrum characteristic of the standard $\Lambda$CDM paradigm by introducing a damping scale $k_b$. Modes with $k>k_b$ enter the horizon before (and may exit during) thermal inflation, so they are strongly influenced by the intermediate stages between the first inflation and the radiation dominated period after the flaton decay. It was shown in \cite{Hong:2015oqa} that the perturbations for $k>k_b$ are strongly suppressed compared with those predicted in the standard $\Lambda$CDM paradigm, so the primordial curvature power spectrum for these models presents a damping at high wavenumbers (small scales). In turn, the matter density perturbations are affected, showing a suppression in the CDM power spectrum at $k> k_b$. Thus, these thermal inflation scenarios belong to the class of non-standard inflation models introduced above. We stress that in non-standard inflation models, the matter power spectrum is naturally suppressed at small scales, without requiring modifications of the standard cold dark matter sector. So, in thermal inflation, DM particles are still massive and non-interacting. Thermal inflation can also produce interesting signatures in CMB observables \cite{Cho:2017zkj} and in the physics of primordial gravitational waves (see e.g. the discussion in \cite{Hong:2017knn}). However, here we will focus only on the effects on the matter fluctuations.

As found in \cite{Hong:2015oqa} the linear matter power spectrum from models of thermal inflation differs from that expected in the standard $\Lambda$CDM by the presence of an enhanced peak in the transfer function at $k\sim k_b$ followed by a damping and oscillations at $k>k_b$. The damping is very similar to that seen in nCDM scenarios. For nCDM models, it is well known that the nonlinear evolution of the Universe at low redshifts transfers power from low to high wavenumbers \cite{Leo:2017wxg,2012MNRAS.421...50V}. The non-linear power spectrum is less affected by the damping, while the halo mass function is more sensitive to the form of the linear power spectrum \cite{Leo:2017wxg}. We expect that this behaviour is true also for thermal inflation. However, the presence of an enhanced peak and oscillations for $k>k_b$ (which are in general not present in simple thermal WDM scenarios, see e.g. \cite{Bode:2000gq,Viel:2005qj}) can potentially introduce new features into  structure formation that deserve to be investigated in detail, and which could potentially leave signatures of thermal inflation in the large-scale structure of the Universe. Here, we investigate the non-linear evolution of structure formation in the thermal inflation scenario described in \cite{Hong:2015oqa} by using high-resolution N-body simulations, highlighting the main differences with respect to the results found in nCDM, other non-standard inflation models and standard $\Lambda$CDM. We note that the impact of thermal inflation on structure formation was addressed recently in \cite{Hong:2017knn}, by e.g. using semi-analytical techniques to calculate dark matter halo abundances. However, we show here that a full study using N-body simulations is necessary to model accurately the non-linear evolution of structure (and to find accurate estimations of the non-linear power spectra and halo abundances at late times). As a second step we compare the N-body results with semi-analytical techniques showing the degree of accuracy of these approaches. 

The paper is structured as follows. In Section 2 we briefly describe the theoretical model of thermal inflation considered here, together with two other models of damped matter fluctuations. In Section 3 we present our N-body simulation set-up. In Section 4 we show our main results for the non-linear power spectra. Section 5 is devoted to the study of halo statistics at $z=0$, while in Section 6 we show the results for halo abundances at higher redshifts. In these sections, we measure the halo mass function from N-body simulations and compare with analytical predictions from a version of the Press-Schechter (PS) approach. Finally, our conclusions are given in Section 7.
\section{Theoretical models}
In this section we briefly describe the model of thermal inflation together with two other damped models: a thermal WDM and a broken scale invariance inflation model. The latter two models are considered because we are interested in quantifying if the results from thermal inflation are, in some way, different  from those found in other damped models. The linear power spectra for all the models considered here are shown in Figure \ref{fig:linearPalla}, while the ratios with respect to $\Lambda$CDM are shown in Figure \ref{fig:linearPallb} (in these figures we show also the power spectra measured from the N-body initial conditions at $z=199$). The $\Lambda$CDM linear power spectrum at $z=199$ is generated with the code {\sc class} \cite{2011arXiv1104.2932L,2011JCAP...09..032L} using the following values for the cosmological parameters: the DM contribution is  $\Omega^0_\mathrm{DM} h^2 = 0.120$, the baryonic contribution is $\Omega_\mathrm{b} h^2 = 0.023$, the dimensionless Hubble constant is $h = 0.6726$, the spectral index of the primordial power spectrum is $n_s =  0.9652$ and the linear rms density fluctuation in a sphere of radius $8$ $h^{-1}$Mpc at $z=0$ is $\sigma_8 = 0.81$. Regarding the other models in Figure \ref{fig:linearPall}, we illustrate how we have generated the associated linear theory power spectra in the subsections below.
\subsection{Thermal inflation}
We consider the model of thermal inflation proposed in \cite{Lyth:1995hj,Lyth:1995ka} and studied in \cite{Hong:2015oqa} from the point of view of the effects on density perturbations. 
This model predicts (at least) two inflationary stages. The universe starts as usual with a standard first (or primordial) inflationary period, which produces nearly scale-invariant perturbations and ends at $t=t_e$. However, since moduli acquire non-null vacuum expectation values (VEV) during the first inflation, this stage is followed by a moduli dominated period (moduli are non-relativistic, so in this stage the Universe is matter dominated), starting at $t=t_a$. In this period a sub-dominant standard radiation component is also present. The moduli dominated era ends when their energy density drops below the constant value $V_0 = V(\phi = 0)$ of the flaton potential, maintained at the origin by thermal effects. At this stage, $t=t_b$, the Universe undergoes a second low-energy inflationary expansion, which dilutes the moduli. Thermal inflation finishes when the thermal bath temperature is not sufficient to hold the flaton at $\phi = 0$. The flaton rapidly rolls to its true minimum, starting to oscillate. At $t=t_c$ a flaton matter dominated period begins and a first-order phase transition converts the flaton energy into standard radiation at $t=t_d$, before BBN. The universe, from this point on, follows the standard history.

Following the convention in \cite{Hong:2015oqa}, we define the characteristic wavenumbers, $k_x \equiv a(t_x) H(t_x)$, with $x=\{a,b,c,d\}$ (where the various times $t_x$ have been introduced in the above paragraph). The numerical values are given e.g. in \cite{Hong:2017knn}. In some thermal inflation scenarios (e.g. multiple thermal inflation \cite{Lyth:1995ka}), the values of $k_a$ and $k_b$ are sufficiently small to be in the range of wavenumbers that are interesting for structure formation. In particular, there are cases when $k_b\ll k_a,k_d$, so the impact of thermal inflation on the curvature power spectrum comes effectively from one parameter, $k_b$, \cite{Hong:2015oqa}. In such cases the curvature power spectrum for thermal inflation can be written as \cite{Hong:2015oqa},

\begin{equation}
P^\mathrm{TI}_\mathcal{R}(k) = P^\mathrm{prim}_\mathcal{R}(k) \, T^2_\mathrm{TI}(k),
\label{eq:PPKTI}
\end{equation}
where $P^\mathrm{prim}_\mathcal{R}(k)$ is the (dimensionless) curvature power spectrum from the first inflationary stage, while $T_\mathrm{TI}(k)$ is the transfer function which contains information about the effects of thermal inflation on the modes with wavenumbers $k>k_b$. The matter power spectrum at a given redshift $z$ is then calculated from the primordial curvature perturbations as $P(k,z) = P^\mathrm{TI}_\mathcal{R}(k) \mathcal{T}^2(k,z)$, where $\mathcal{T}(k,z)$ is the transfer function that characterises the evolution after the flaton decay. 

$P^\mathrm{prim}_\mathcal{R}$ takes the approximate form (as calculated in \cite{Cho:2017zkj}),
\begin{equation}
P^\mathrm{prim}_\mathcal{R} \simeq A_* \left(1 - \frac{1}{{N}_*} \ln\left(\frac{k}{k_*}\right) \right)^{(1-n_*){N}_*},
\end{equation}
where $A_*$ is the amplitude at the pivotal scale $k_*$ and $N_*$ is 
\begin{equation}
N_* \equiv \ln\left(\frac{k_e}{k_*}\right).
\end{equation}
$N_*$ is uncertain mainly because of the unknown phases before the moduli domination epoch (see \cite{Cho:2017zkj} for the interval of possible values that can be taken by $N_*$).

The thermal inflation transfer function $T_\mathrm{TI}(k)$, takes the analytical form \cite{Hong:2015oqa},
\begin{equation}
\begin{split}
T_\mathrm{TI} (k) &= \cos\left[\left(\frac{k}{k_b}\right)\,\int^\infty_0 \frac{d\alpha}{\sqrt{\alpha(2+\alpha^3)}}\right]\\&+6\left(\frac{k}{k_b}\right)\int^\infty_0 \frac{d\gamma}{\gamma^3}\int^\infty_0 d\beta \left(\frac{\beta}{2+\beta^3}\right)^{3/2} \sin\left[\left(\frac{k}{k_b}\right)\,\int^\infty_\gamma  \frac{d\alpha}{\sqrt{\alpha(2+\alpha^3)}}\right].\end{split}
\end{equation}
The above expression is unity for $k\ll k_b$, and corresponds to an enhancement of $\sim 20\%$  around $k\simeq 1.13\, k_b$, while for $k\gg k_b$ the transfer function oscillates around zero as $T_\mathrm{TI}(k)\simeq -\cos(2.23 \, k/k_b)/5$.  

The primordial curvature power spectrum for the standard $\Lambda$CDM paradigm is
\begin{equation}
P^\mathrm{prim,s}_\mathcal{R}(k) = A_*\left(\frac{k}{k_*}\right)^{n_*-1},
\label{eq:PRstandard}
\end{equation}
where the pivotal wavenumber is $k_* =0.05\,\mathrm{Mpc}^{-1}$.
Three differences arise when comparing the power spectrum from thermal inflation (eq. (\ref{eq:PPKTI})) with that from the standard paradigm (eq. (\ref{eq:PRstandard})):
\begin{itemize}
\item a small change in the $P^\mathrm{prim}_\mathcal{R}$ with respect to $P^\mathrm{prim,s}_\mathcal{R}$ due to $N_*$ (this difference is $k$-dependent). However, as shown in \cite{Hong:2017knn} the difference due to the choice of $N_*$ is negligible (if compared to the other effects listed below) at $k\lesssim k_b$ for $k_b\geq 1 \,\mathrm{Mpc}^{-1}$, while the range of wavenumbers really affected by $N_*$ are those ($k\gg k_b$, oscillation regime) which are already extremely damped by $T^2_\mathrm{TI}$ (see Figure 3 or 4 in \cite{Hong:2017knn}). So, instead of fixing a value of $N_*$, from here on we will consider $P^\mathrm{prim}_\mathcal{R}(k) = P^\mathrm{prim,s}_\mathcal{R}(k)$;
\item an enhancement in the power amplitude at $k\sim k_b$, and
\item a strong damping for $k> 3k_b$, with an oscillatory pattern in the power spectrum of thermal inflation ($T^2_\mathrm{TI}$ oscillates around $1/50$ at large wavenumbers). 
\end{itemize}

\begin{figure}[t]
\advance\leftskip-.4cm
\advance\rightskip-3cm
\subfigure[][]
{\includegraphics[width=.55\textwidth]{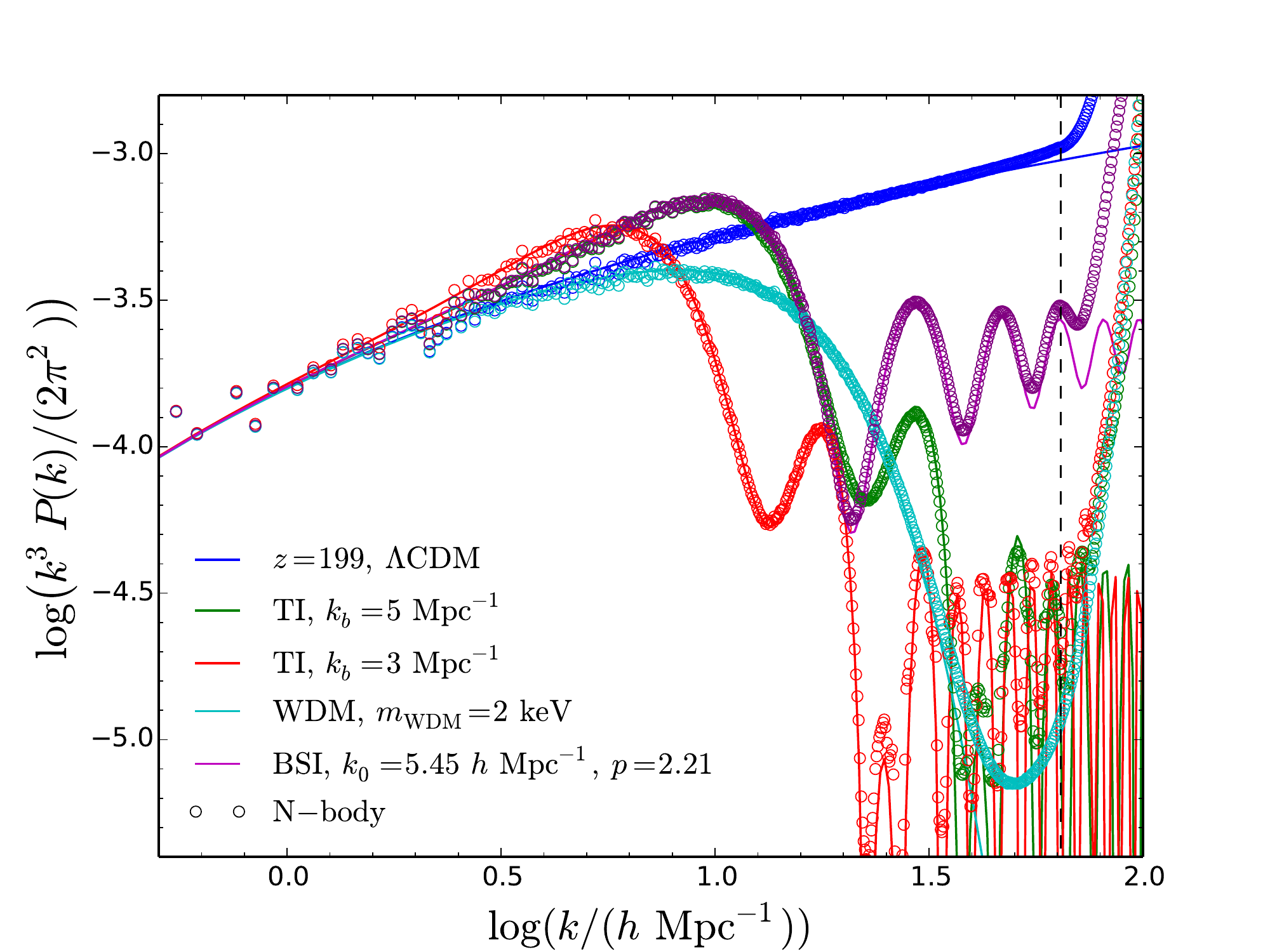}\label{fig:linearPalla}}\hspace{-2.\baselineskip}
\subfigure[][]
{\includegraphics[width=.55\textwidth]{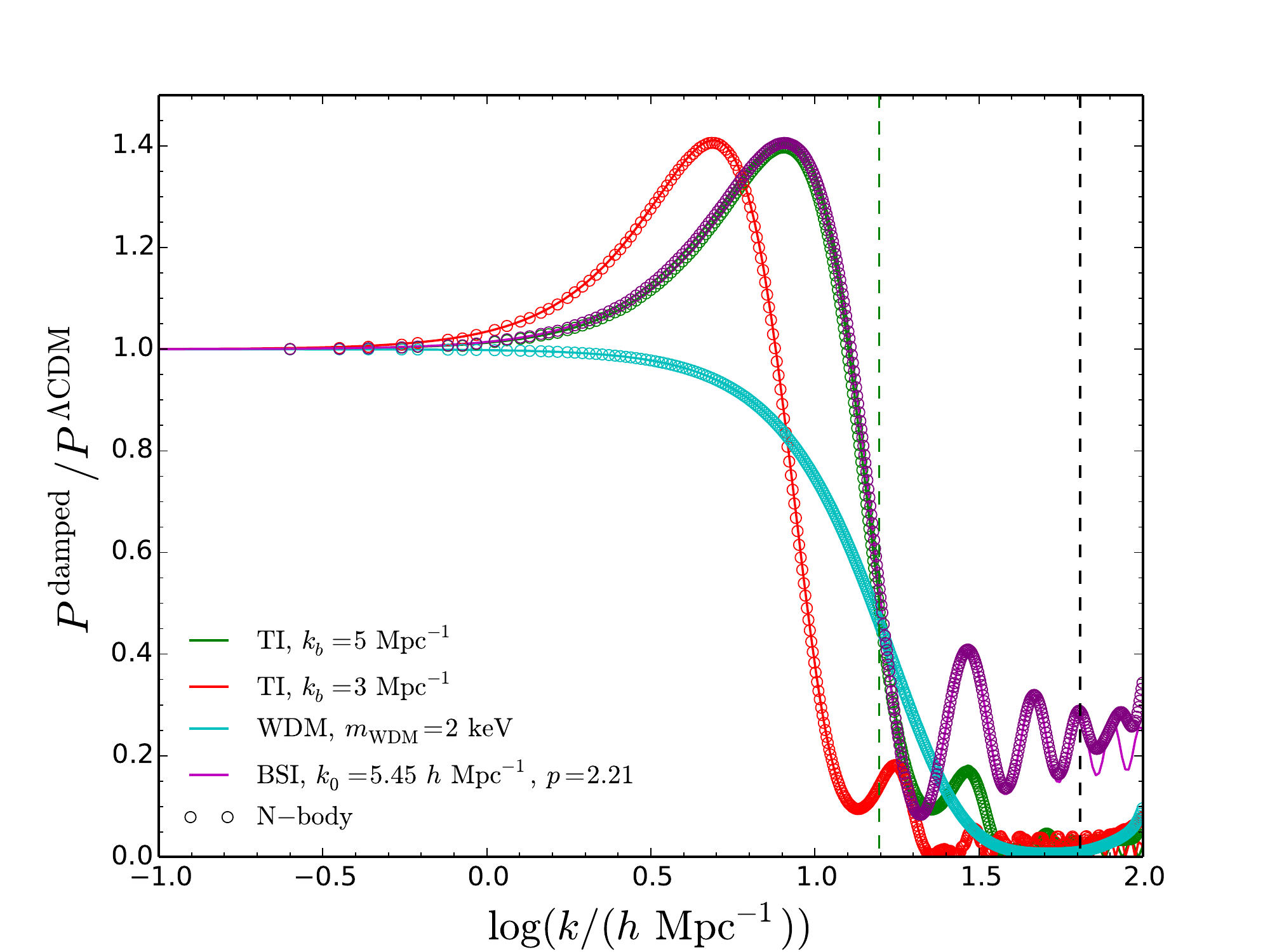}\label{fig:linearPallb}}\\
\caption{(a) Matter power spectra generated at $z=199$ for different models as labelled. (b) Ratios of the linear damped power spectra relative to that from standard $\Lambda$CDM. Solid lines show the linear theory power spectra, while symbols represent the power spectra measured from the N-body ICs. The black vertical dashed line indicates the Nyquist frequency of the simulations.  The green vertical dashed line in panel (b) shows the position of the half-mode wavenumber $k_{1/2,\, \mathrm{TI5}}$ for the thermal inflation matter power spectrum with $k_b=5\,\mathrm{Mpc}^{-1}$.}
\label{fig:linearPall}

\end{figure}

To calculate the matter power spectrum in thermal inflation, we have used the {\sc class} code \cite{2011arXiv1104.2932L,2011JCAP...09..032L}, providing as input the primordial curvature power spectrum for thermal inflation. The matter power spectra $P(k)$ at $z=199$ are shown in Figure \ref{fig:linearPall}  for the two values of the characteristic wavenumber $k_b=5$ and $3\,\mathrm{Mpc}^{-1}$ considered in this analysis, together with that from standard $\Lambda$CDM and two other damped models (see below) for comparison. We choose these two values of $k_b$ because they produce a sufficient reduction of the number of haloes with mass $M_\mathrm{halo}<10^{9}\,h^{-1}\,\mathrm{M}_\odot$ (as we will see in the next sections) to be considered as possible solutions to the missing satellite problem. Larger values of $k_b$ give matter power spectra that are less suppressed and so are very similar to standard $\Lambda$CDM at the scales of interest in our analysis, and as pointed out in \cite{Hong:2015oqa,Cho:2017zkj} only $k_b\gtrsim 1 \,\mathrm{Mpc}^{-1}$ is allowed by CMB constraints.  As can be seen in Figure \ref{fig:linearPallb}, the enhancement at $k\sim k_b$ and the damping at larger wavenumbers influence significantly the shape of the matter power spectrum of thermal inflation when comparing with that from standard $\Lambda$CDM, so we will focus on the effect of these two properties on structure formation. We also note that the matter power spectra from thermal inflation are significantly different, in general, from those expected from nCDM (see e.g. the thermal WDM power spectrum in Figure \ref{fig:linearPall} and the discussion below), because of this enhancement in power at $k\sim k_b$ and the presence of oscillations (although some nCDM scenarios such as axion-like DM \cite{Marsh:2015xka} or interacting DM \cite{Boehm:2004th,Boehm:2014vja,Schewtschenko:2014fca,Spergel:1999mh} also display oscillations in the matter $P(k)$).

\subsection{Thermal WDM}
Since we are interested in how  structure formation in thermal inflation differs from that expected in a nCDM model, we consider here the simple thermal WDM scenario. For this model, the transfer function, $T(k)=\sqrt{P^\mathrm{WDM}/P^\mathrm{CDM}}$, takes the approximate form \cite{Bode:2000gq,Viel:2005qj},
\begin{equation}
T(k) = \left(1+(\alpha k)^\beta\right)^\gamma,
\end{equation}
where
\begin{equation}
\alpha = a  \left(\frac{\Omega^0_\mathrm{WDM}}{0.25}\right)^b  \left(\frac{h}{0.7}\right)^c  \left(\frac{m_\mathrm{WDM}}{\text{keV}}\right)^d, \quad\beta = 2\nu,\quad \gamma = -5/\nu,\end{equation}
and $a = 0.049,\, b = 0.11, \, c = 1.22, \, d = - 1.11, \, \nu = 1.12$, as computed in \cite{Viel:2005qj}.
We choose the WDM particle mass $m_\mathrm{WDM}=2\,\mathrm{keV}$ because the corresponding half-mode wavenumber\footnote{We define the half-mode wavenumber $k_{1/2,\,\mathrm{damped}}$ for a given damped model as the wavenumber at which the ratio $P^\mathrm{damped}/P^\mathrm{\Lambda CDM}$ between the linear damped and standard $\Lambda$CDM power spectrum is equal to $1/2$.} ($k_{1/2,\mathrm{WDM2}}$)  is roughly equal to the half-mode wavenumber ($k_{1/2,\mathrm{TI5}}$) for the thermal inflation model with $k_b=5\mathrm{Mpc}^{-1}$ (see Figure \ref{fig:linearPallb}), so these two models are directly comparable. The WDM power spectrum considered here is shown in Figure  \ref{fig:linearPall}.

\subsection{BSI inflation}
We also compare thermal inflation with another non-standard inflation model, which hereafter we call \emph{broken scale invariance} inflation or BSI. Inspired by the scenario proposed in \cite{Starobinsky:1992ts,Lesgourgues:1997en} (which was studied as a viable solution of the small-scale crisis in \cite{Kamionkowski:1999vp}), we consider a model where the primordial curvature spectrum takes the form $P^\mathrm{prim}_\mathcal{R}(k) = P^\mathrm{prim,s}_\mathcal{R}(k)\, T^2_\mathrm{BSI}(k)$, where $P^\mathrm{prim,s}_\mathcal{R}$ is the standard $\Lambda$CDM primordial spectrum (see Eq. (\ref{eq:PRstandard})), while $T^2_\mathrm{BSI}$ is given by \cite{Starobinsky:1992ts} 
\begin{equation}
\begin{split}
&\frac{T^2_\mathrm{BSI}(k)}{\mathcal{N}_{\mathrm{BSI}}} = 1-3 (p-1) \,\left(\frac{k_0}{k}\right)\, \left[\left(1-\frac{k^2_0}{k^2}\right)  \sin\left(\frac{2\,k}{k_0}\right)+ \left(\frac{2\,k_0}{k}\right)  \cos\left(\frac{2\,k}{k_0}\right)\right]\\
&+\, \frac{9}{2} \, (p-1)^2 \, \left(\frac{k_0}{k}\right)^2 \, \left(1+\frac{k^2_0}{k^2}\right) \, \left[\left(1+\frac{k^2_0}{k^2}\right) + \left(1+\frac{k^2_0}{k^2}\right) \, \cos\left(\frac{2\,k}{k_0}\right) - \left(\frac{2\,k_0}{k}\right)  \sin\left(\frac{2\,k}{k_0}\right)\right],
\end{split}
\end{equation}
where $k_0$ is the wavenumber above which the power spectrum breaks its scale-invariance, $p$ quantifies the power suppression\footnote{Note that if $p>1$ the power is suppressed at high wavenumbers, while if $p<1$ the power is enhanced at high wavenumbers.} and the normalisation $\mathcal{N}_\mathrm{BSI}$ is chosen such that $T^2_\mathrm{BSI}(k) = 1$ for $k\ll k_0$, so at small wavenumbers the BSI power spectrum is equal to that in the standard paradigm. We note that this transfer function has an enhanced peak at $k\sim k_0$, whose amplitude depends on $p$. As we did for the case of thermal inflation, we have modified {\sc class} providing the BSI primordial power spectrum as input. In this way, {\sc class} can calculate the linear theory matter power spectrum for BSI. We choose the free parameters, $\{k_0,p\}$, such that the linear matter power spectrum for BSI has the enhanced peak in the same position and with the same amplitude as the linear $P(k)$ for the case of thermal inflation with $k_b=5\,\mathrm{Mpc}^{-1}$. We find that for $k_0=5.45\, h\,\mathrm{Mpc}^{-1}$ and $p=2.21$, the enhanced peak in BSI linear matter power spectrum is roughly equal to that of thermal inflation with $k_b=5\,\mathrm{Mpc}^{-1}$. The linear power spectrum for the BSI model is shown in Figure \ref{fig:linearPall}. From Figure \ref{fig:linearPallb} the choice of this BSI model to compare with thermal inflation with $k_b=5\,\mathrm{Mpc}^{-1}$ is clear. Indeed, the linear matter power spectrum of BSI is very similar to that of thermal inflation with $k_b=5\,\mathrm{Mpc}^{-1}$ up to $k\sim 20 \,h\,\mathrm{Mpc}^{-1}$. For larger wavenumbers, the thermal inflation transfer function oscillates around zero, while the BSI $T_\mathrm{BSI}(k)$ oscillates around a constant non-zero value\footnote{Apart from the enhanced peaks, the differences in the transfer functions at high wavenumbers between BSI and thermal inflation are very similar to those between mixed DM (see e.g. \cite{Boyarsky:2008xj}) and pure WDM. Indeed, pure WDM models have vanishing transfer functions at high wavenumbers (see e.g. the thermal WDM power spectrum in  Figure \ref{fig:linearPallb}). While, in mixed DM (where the dark matter content is made up of a mixture of warm and cold DM) the transfer function reaches a constant non-zero value at high wavenumbers since CDM fluctuations are present on small scales \cite{Boyarsky:2008xj}.}. However, we note that the power spectrum depends on the squared transfer function. In the case of thermal inflation $T^2_\mathrm{TI}(k)$ oscillates around $1/50$ at high wavenumbers (see Section 2.1) while for BSI $T^2_\mathrm{BSI}(k)\sim 0.22$. The thermal inflation linear power spectrum is then more suppressed at large wavenumbers than that from BSI.
\section{N-body simulations}
The linear matter power spectra shown as solid lines in Figure \ref{fig:linearPall} are used to generate the initial conditions (ICs) for N-body simulations,  using the second-order Lagrangian perturbation code 2LPTic \cite{Crocce:2006ve}. The initial redshift is chosen to be $z=199$ to ensure that all the modes probed in our analysis are in the linear regime. The simulations are performed in a cubic box of comoving length $L=25\,h^{-1}\,\mathrm{Mpc}$ using $N=512^3$ particles (the simulation particle mass is $\sim10^7\,h^{-1}\,\mathrm{M}_\odot$). We choose this pair of $\{N,L\}$ because we want to resolve structures on scales near the cut-off of the thermal inflation linear power spectra (see Figure \ref{fig:linearPall} and Section 4 below). We have extensively tested the accuracy of simulations with this choice of parameters against possible numerical effects (see Appendix A and our previous work on thermal WDM models \cite{Leo:2017zff}). The Nyquist frequency of a simulation is $k_\mathrm{Ny}\equiv \pi \,(N^{1/3}/L)$ (this specifies the value up to which we can trust the $P(k)$). We evolve the ICs to $z=0$ using the publicly-available tree-PM code Gadget2 \cite{Springel:2005mi}. The gravitational softening length is chosen to be $1/40$-th of the mean inter-particle separation, $L/N^{1/3}$.

\section{Matter power spectra}

Below we present our results for the matter power spectra measured from the simulations. We show our results for seven  redshifts $z=199$ (initial conditions) and $z=19,9,5,3,1,0$. The matter power spectrum is measured from the Gadget2 snapshots using a code based on the cloud-in-cell mass assignment scheme. 

\begin{figure}[t!]
\advance\leftskip-.4cm
\advance\rightskip-3cm
\subfigure[][$k_b=5\,\mathrm{Mpc}^{-1}$]
{\includegraphics[width=.55\textwidth]{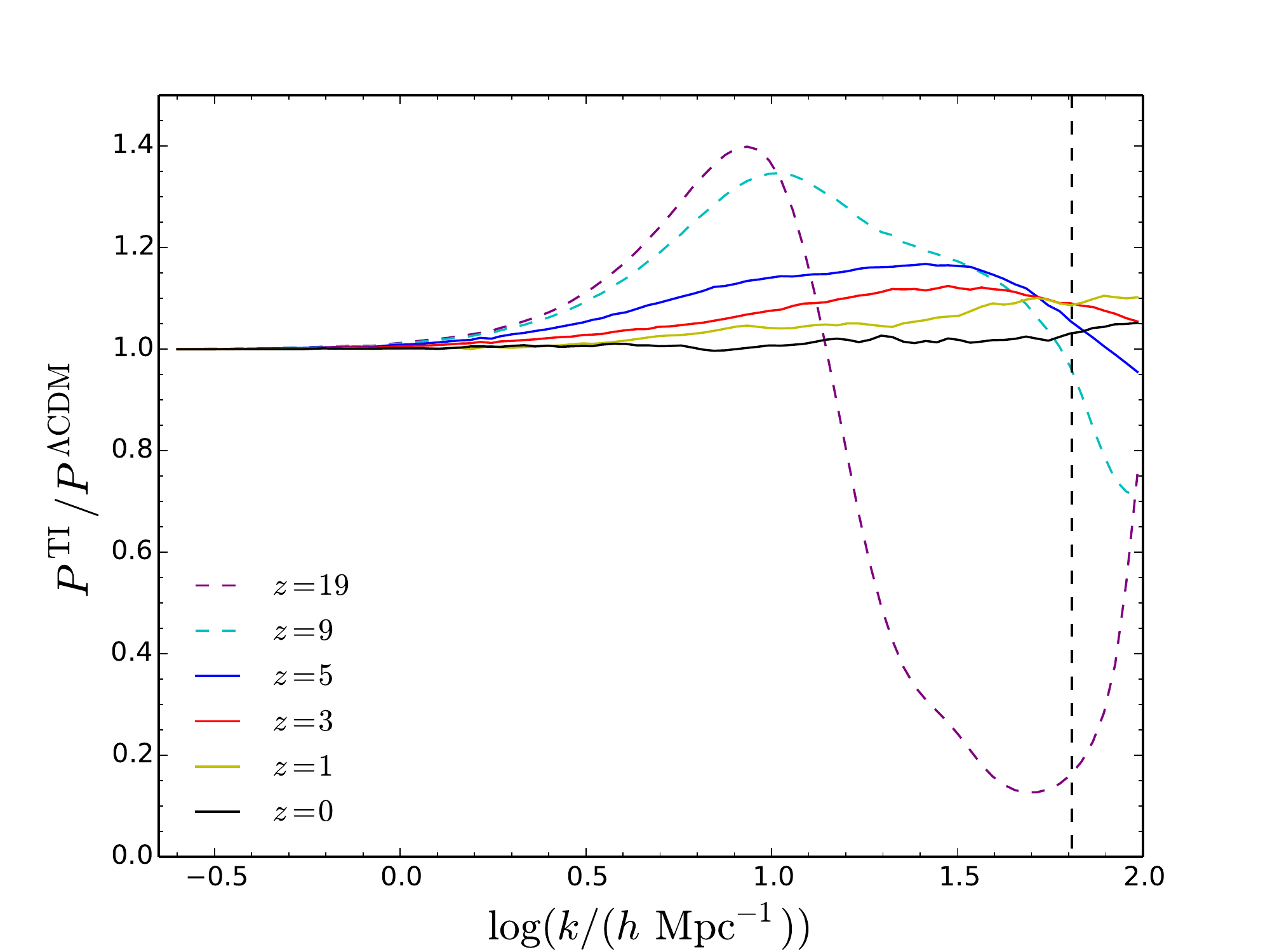}\label{fig:EvolN-bodya}}\hspace{-2.\baselineskip}
\subfigure[][$k_b=3\,\mathrm{Mpc}^{-1}$]
{\includegraphics[width=.55\textwidth]{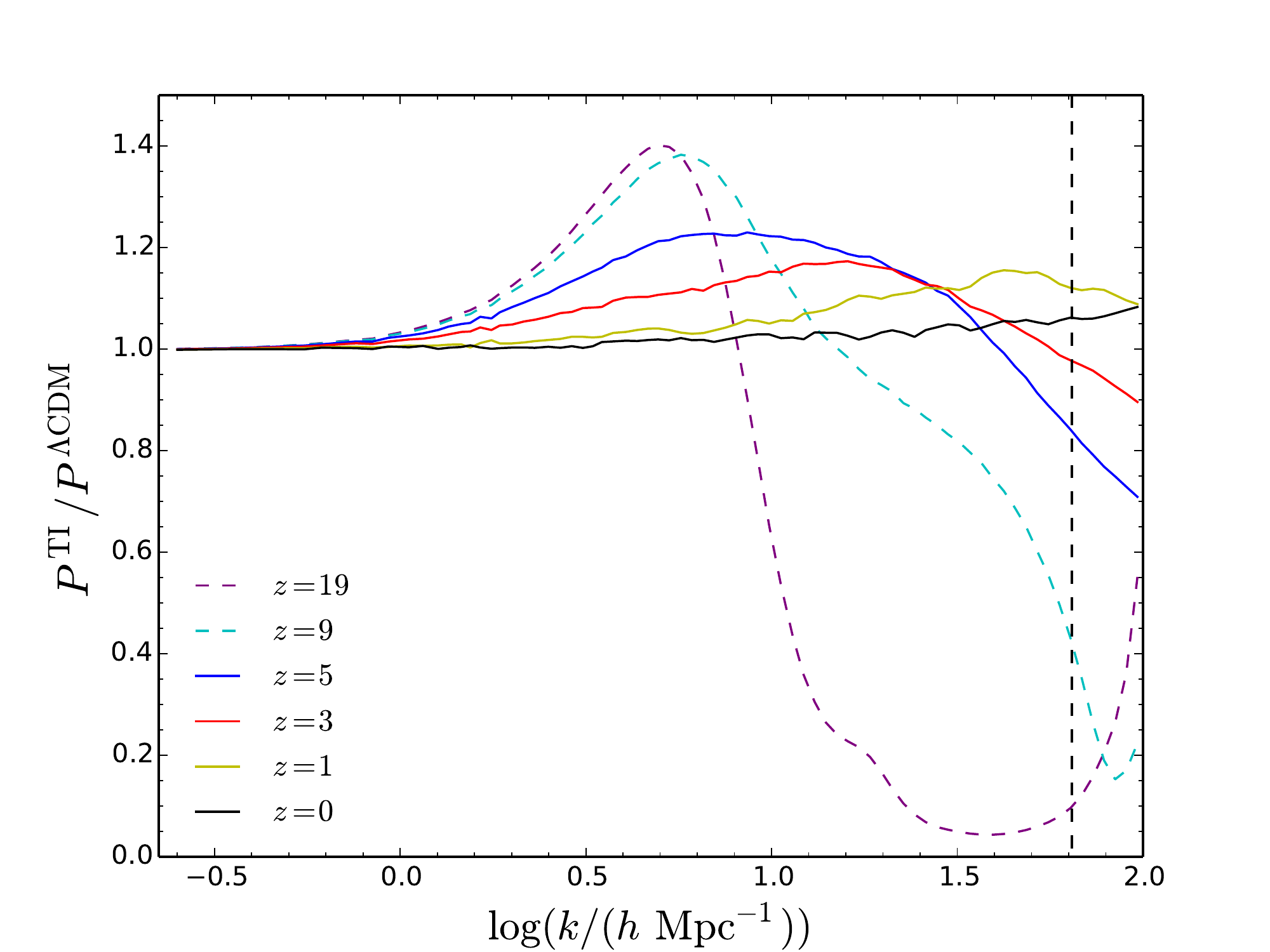}\label{fig:EvolN-bodyb}}\\
\caption{Ratios of the matter power spectra measured from N-body simulations of thermal inflation respect to those measured from standard $\Lambda$CDM simulations at various redshifts (as labelled).  Panel (a) shows the results for thermal inflation with $k_b=5\,\mathrm{Mpc}^{-1}$, while the results for $k_b=3\,\mathrm{Mpc}^{-1}$ are shown in panel (b). The black vertical dashed line indicates the Nyquist frequency of the simulations.}
\label{fig:EvolN-body}

\end{figure}

\emph{Initial $P(k)$} -- The matter power spectra measured from the ICs for all the models described in Section 2 are shown as symbols in Figure \ref{fig:linearPalla}. These are presented normalised as $\Delta^2(k)\equiv k^3 P(k)/(2\pi^2)$. The ratios $P^\mathrm{damped}/P^\mathrm{\Lambda CDM}$ of the damped power spectra with respect to that from $\Lambda$CDM are displayed in Figure \ref{fig:linearPallb} instead. As shown in these figures, the ICs resolve well the cut-off region for all the power spectra considered in our analysis. It is interesting to note how well the ICs capture the enhanced peak at $k\sim k_b$  and the oscillatory behaviour at $k> k_{b}$ in the case of the thermal inflation models. Below we will see how the enhancement in the thermal inflation $P(k)$ changes the non-linear power spectrum and if the oscillatory pattern at high wavenumbers survives non-linear evolution.

\emph{Evolved $P(k)$} -- The matter power spectra  at late times for the two thermal inflation models are shown in Figure \ref{fig:EvolN-body}. First, we note that the oscillations at high wavenumbers do not survive the non-linear evolution and they are erased at low redshifts, in agreement with what we found in \cite{Leo:2017wxg} for an oscillatory power spectrum. Second, the enhanced peak in the linear power spectra is progressively shifted to higher wavenumbers in the non-linear regime, while the peak height is reduced. By $z=0$, the thermal inflation $P(k)$ are very similar to the $P(k)$ of the standard $\Lambda$CDM at all the wavenumbers probed by our simulations. We note that in the linear regime, the thermal inflation power spectra were extremely suppressed at $k>3\,k_b$ (see Figure \ref{fig:linearPallb}). However, due to the shift of the peak position to large wavenumbers, the non-linear power spectra for these models show, in general, more power at $k>3\,k_b$ with respect to $\Lambda$CDM. This is true for both the thermal inflation power spectra considered here. For example, at $z=0$, for the model with $k_b = 5\,\mathrm{Mpc}^{-1}$ we have $P^\mathrm{TI}/P^\mathrm{\Lambda CDM}\simeq 1.02$, while for the thermal inflation with $k_b = 3\,\mathrm{Mpc}^{-1}$ $P^\mathrm{TI}/P^\mathrm{\Lambda CDM}\simeq 1.05$ at wavenumber $k\sim 32\,h\,\mathrm{Mpc}^{-1}$. In the linear regime, both thermal inflation power spectra where suppressed by $\sim1/50$ at $k\sim 32\,h\,\mathrm{Mpc}^{-1}$ with respect to $\Lambda$CDM.

\begin{figure}[t]
\advance\leftskip-.4cm
\advance\rightskip-3cm
\subfigure[][$z=5$]
{\includegraphics[width=.55\textwidth]{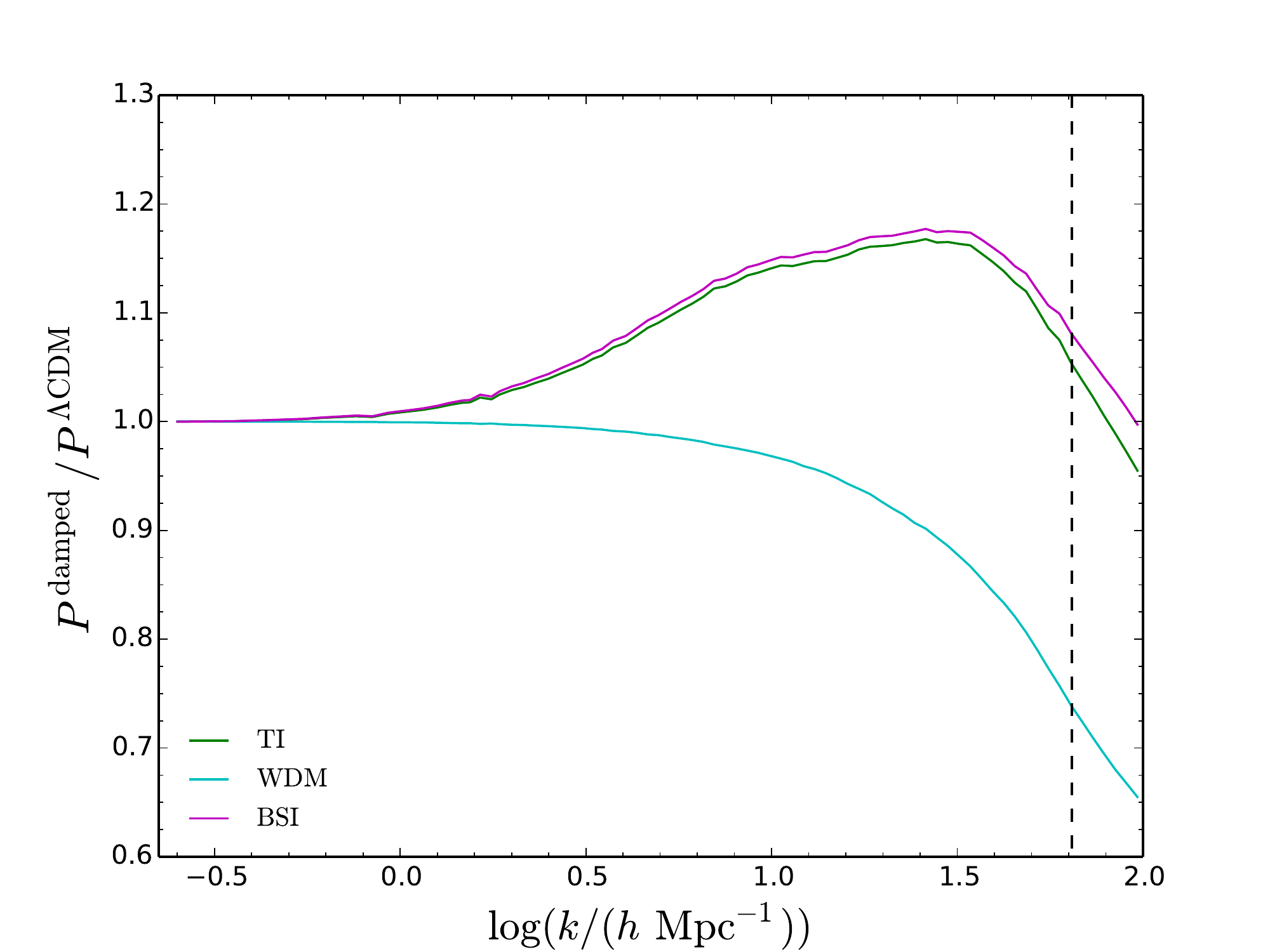}}\hspace{-2.\baselineskip}
\subfigure[][$z=3$]
{\includegraphics[width=.55\textwidth]{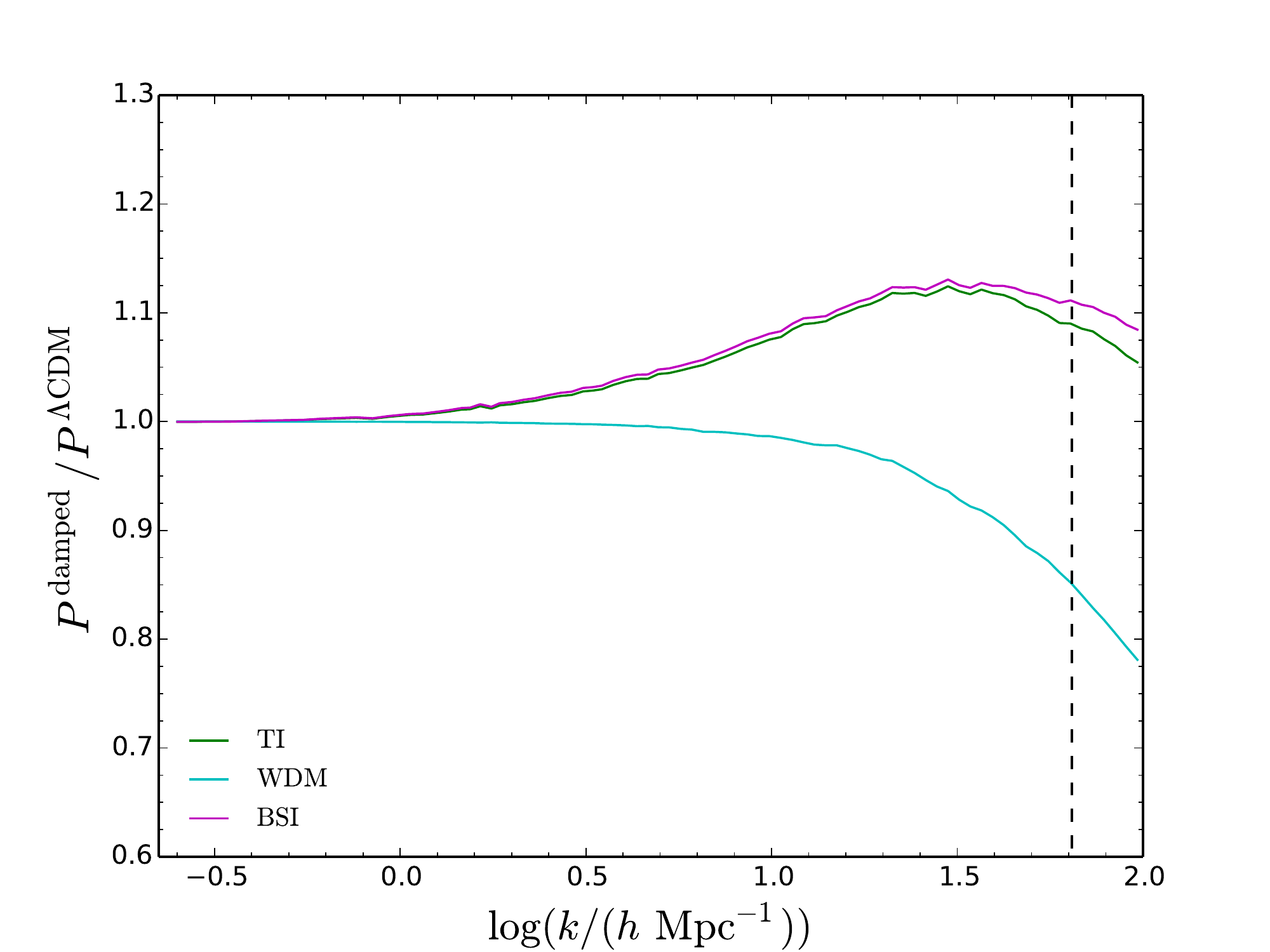}}\\
\subfigure[][$z=1$]
{\includegraphics[width=.55\textwidth]{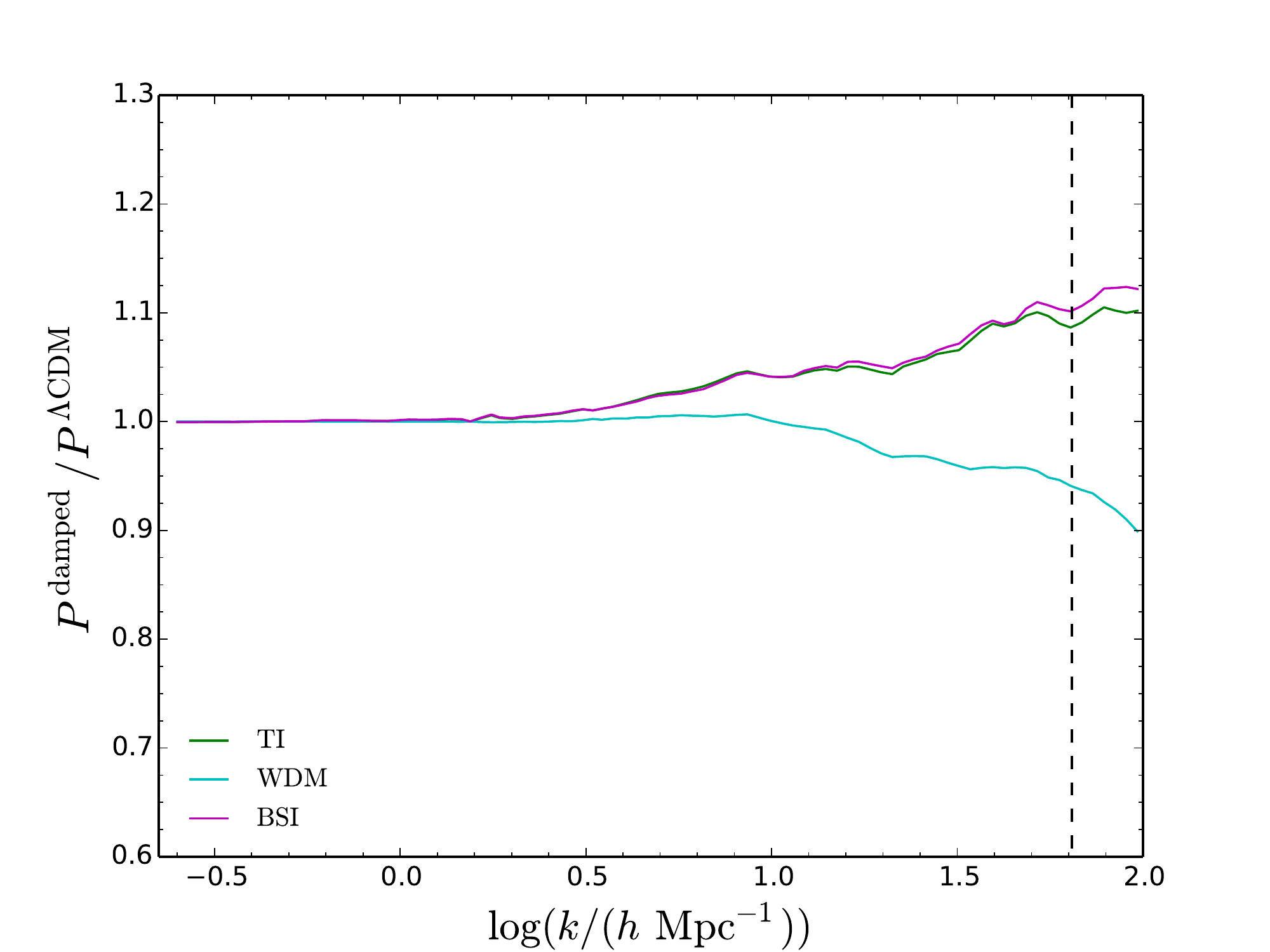}}\hspace{-2.\baselineskip}
\subfigure[][$z=0$]
{\includegraphics[width=.55\textwidth]{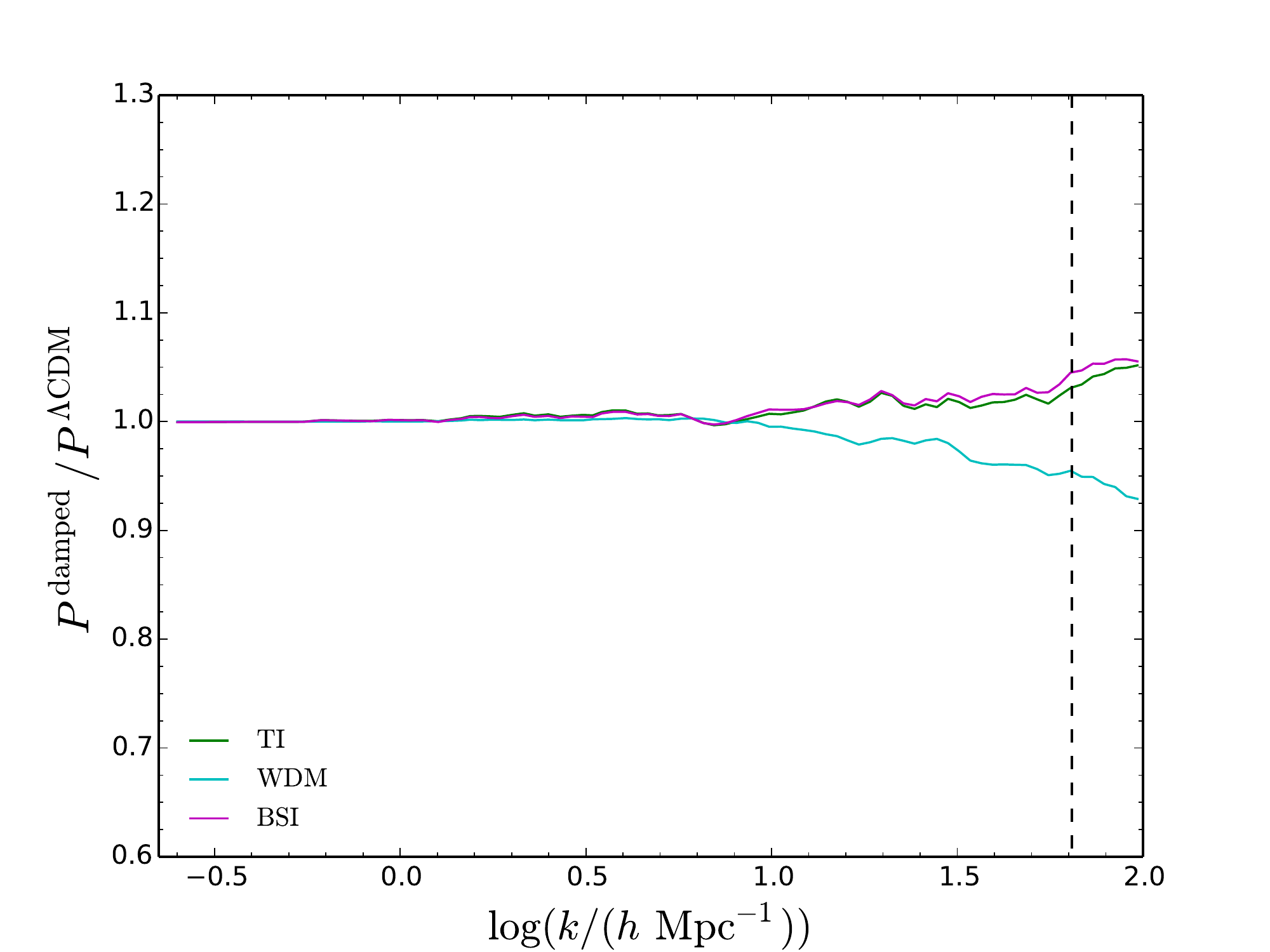}}\\
\caption{Ratios of the matter power spectra for thermal inflation with $k_b=5\,\mathrm{Mpc}^{-1}$ (green), thermal WDM (cyan) and BSI (magenta) with respect to those from standard $\Lambda$CDM measured from N-body simulations at redshifts $z = 5,3,1,0$ as labelled. The black vertical dashed line indicates the Nyquist frequency of the simulations.}
\label{fig:WDMN-body}

\end{figure}

\emph{Comparison with WDM and BSI} -- We have also compared the results from thermal inflation with $k_b=5\,\mathrm{Mpc}^{-1}$ with those from the WDM and BSI models described in Section 2.  The non-linear power spectra measured from N-body simulations for these models (shown as ratios to standard $\Lambda$CDM) are given in Figure \ref{fig:WDMN-body}. For thermal WDM (but see also \cite{Leo:2017wxg} for more results about nCDM-like models), although the non-linear evolution transfers the power from small to large scales, the non-linear power spectra at low $z$ always have less power than the standard $\Lambda$CDM $P(k)$, in contrast with what we find for thermal inflationary models (see the discussion in \emph{Evolved P(k)} above). In the case of the BSI model, the non-linear power spectrum behaviour follows  that of  thermal inflation. Indeed, as can be seen from Figure \ref{fig:WDMN-body} the enhanced peak is reduced in magnitude and shifted by the same amounts for both thermal inflation and BSI. In general, the matter power spectra at low redshift from thermal inflation display less power than BSI. However, the differences in the non-linear matter power spectra at $k\gg k_0,k_b$ between thermal inflation and BSI are appreciably less than those in the linear theory power spectra on the same scales. For example,  the ratio between thermal inflation and BSI power spectrum at $z=199$ (linear theory) is small, $\sim 0.09$ at $k\sim k_\mathrm{Ny}$, while the ratio between the two non-linear power spectra at $z=0$ is  $\sim 0.98$ for wavenumbers near the Nyquist frequency. The non-linear transfer of power from large to small scales has reduced the initial linear theory difference between these two models by a factor of $10$.

In conclusion, non-linear matter power spectra are then a blunt tool to distinguish the effects of thermal inflation or BSI (in line with what we found for nCDM models in \cite{Leo:2017wxg}). We will see below that halo statistics are more sensitive to the shape of the linear $P(k)$. We stress, however, that the results regarding the non-linear $P(k)$ have not appeared in the literature before, since previous studies on damping models (including our study \cite{Leo:2017wxg}) have always focused on the damping features of the matter power spectra. In thermal and BSI inflation, we find that the presence of an enhanced peak in the linear power spectrum affects substantially the behaviour of the non-linear power spectra at small scales. The non-linear power spectra in these models are then different from those found in nCDM scenarios, particularly at high redshifts.
\section{Halo statistics at $z=0$}
In this Section we explore whether counting the number of haloes of different masses can discriminate between thermal inflation and the standard paradigm. We will see also if the thermal inflation models predict a different halo mass function than thermal WDM and BSI. In this Section we focus on results at $z=0$, we discuss the halo mass function at high redshifts in the next Section.
\subsection{Measured halo mass function}
Regardless of the nature of the process producing damping of matter fluctuations on small scales, one common impact of these models on structure formation is a reduction in halo abundance at low masses (see e.g. \cite{Bode:2000gq,Wang:2007he,Lovell:2013ola,Schewtschenko:2014fca,2012MNRAS.424..684S,Power:2013rpw,Power:2016usj, Schneider:2013ria, Schneider:2014rda,2012MNRAS.420.2318L,Bose:2015mga}), offering a possible solution to the missing satellite problem. Since thermal inflation models are characterised by damping in the linear matter power spectrum at high wavenumbers (as seen above), we expect a similar reduction in the number of low-mass haloes with respect to standard $\Lambda$CDM. However, since thermal inflation power spectra are characterised by enhanced peaks, it is also possible to find such features imposed on the halo mass function\footnote{These features  in thermal inflation cosmologies have been found in \cite{Hong:2017knn} by inferring the halo mass function by using the analytical PS approach. We review this approach in the next subsection by showing that using a spherical top-hat filter the PS analytical predictions over-estimate the small mass halo abundance when compared with N-body simulations.}. In this subsection we show the halo mass function measured  from the N-body simulations at $z=0$ (for halo mass functions at high redshifts see Section 6).  To extract the halo properties from simulation outputs we use  the code {\sc rockstar} which is a phase-space friends-of-friends halo finder \cite{2013ApJ...762..109B}. As a definition of the halo mass, we use the mass, $M_{200}$, contained in a sphere of radius $r_{200}$, within which the average density is $200$ times the critical density of the universe at the specified redshift. Below the (differential) halo mass function is presented as $F(M_{200},z)=dn/d\log(M_{200})$,  where $n$ is the number density of haloes with mass $M_{200}$. We assume a minimum of $50$ simulation particles in a halo, so the minimum halo mass is $\sim5\times10^{8}\, h^{-1}\,\mathrm{M}_\odot$. We note also that the volume of our simulations is too small to provide a statistically robust sample of haloes with masses $M_{200}>10^{12}\,h^{-1}\,\mathrm{M}_\odot$. This means that for the most massive haloes in our simulations the results are influenced  by large Poisson fluctuations.

N-body simulations of damped models display the effects of artificial fragmentation, with regularly-spaced clumps (spurious haloes) along filaments, the distance between which reflects the initial inter-particle separation  \cite{Bode:2000gq,Wang:2007he,Lovell:2013ola,Schewtschenko:2014fca,2012MNRAS.424..684S,Power:2013rpw,Power:2016usj, Schneider:2013ria, Schneider:2014rda,2012MNRAS.420.2318L,Bose:2015mga}. If not carefully  identified and removed, spurious haloes can influence dramatically the measured halo abundances at small masses. An estimate of the mass below which spurious haloes are likely to be found was proposed in \cite{Wang:2007he}, \begin{equation}
M_\mathrm{lim} = 10.1 \,\bar{\rho} \, d \, k^{-2}_\mathrm{peak},
\end{equation}
where $\bar{\rho}$ is the mean density of the Universe, $d$ is the mean inter-particle separation in the simulation and $k_\mathrm{peak}$ is the wavenumber at which the dimensionless power spectrum, $\Delta^2(k)= k^3 P(k)/(2\pi^2)$, reaches its maximum\footnote{Note that in N-body simulations of WDM models, if thermal velocities are added to the gravitational-induced velocities of the computational particles, $M_\mathrm{lim}$ is shifted to higher masses due to the extra noise introduced in the simulations because of thermal velocities \cite{Leo:2017zff}.}. 

\begin{figure}[t]
\advance\leftskip-.4cm
\advance\rightskip-3cm
\quad\quad\quad\quad\quad\quad\quad\quad\quad\quad\subfigure[][]
{\includegraphics[width=.55\textwidth]{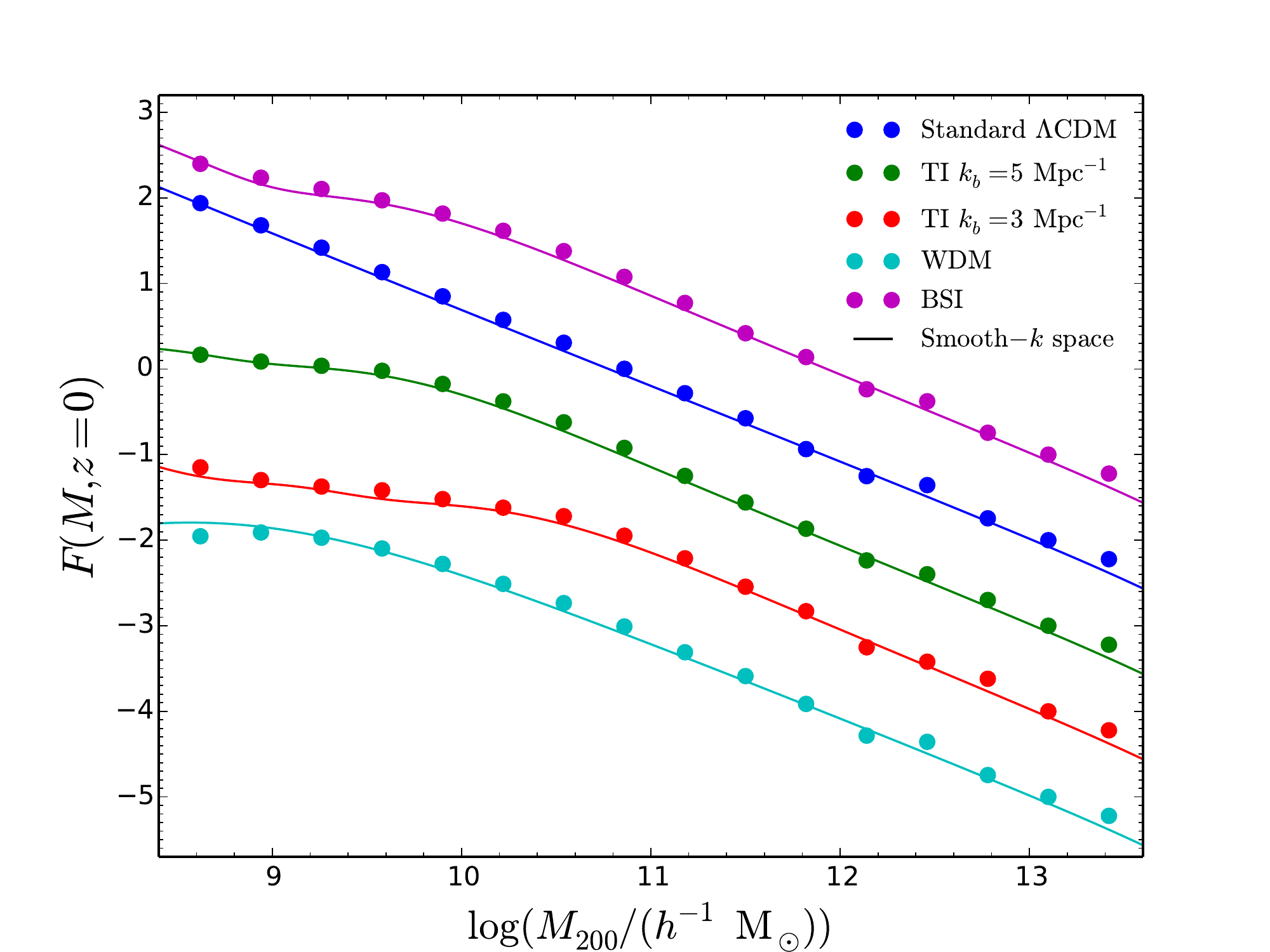}\label{fig:allhmfa}}\\
\subfigure[][Uncleaned catalogues]
{\includegraphics[width=.55\textwidth]{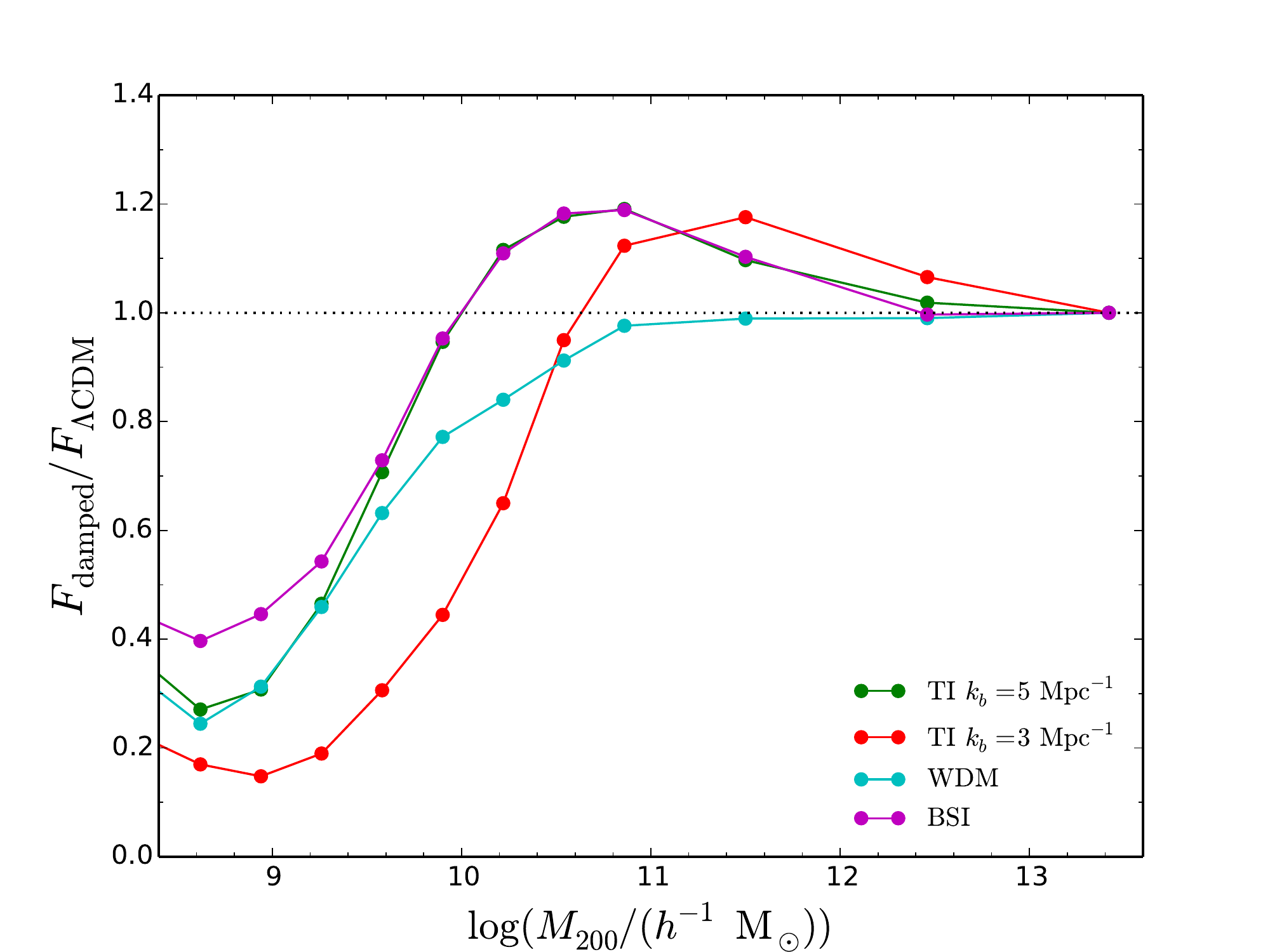}\label{fig:allhmfb}}\hspace{-1.5\baselineskip}
\subfigure[][Cleaned catalogues]
{\includegraphics[width=.55\textwidth]{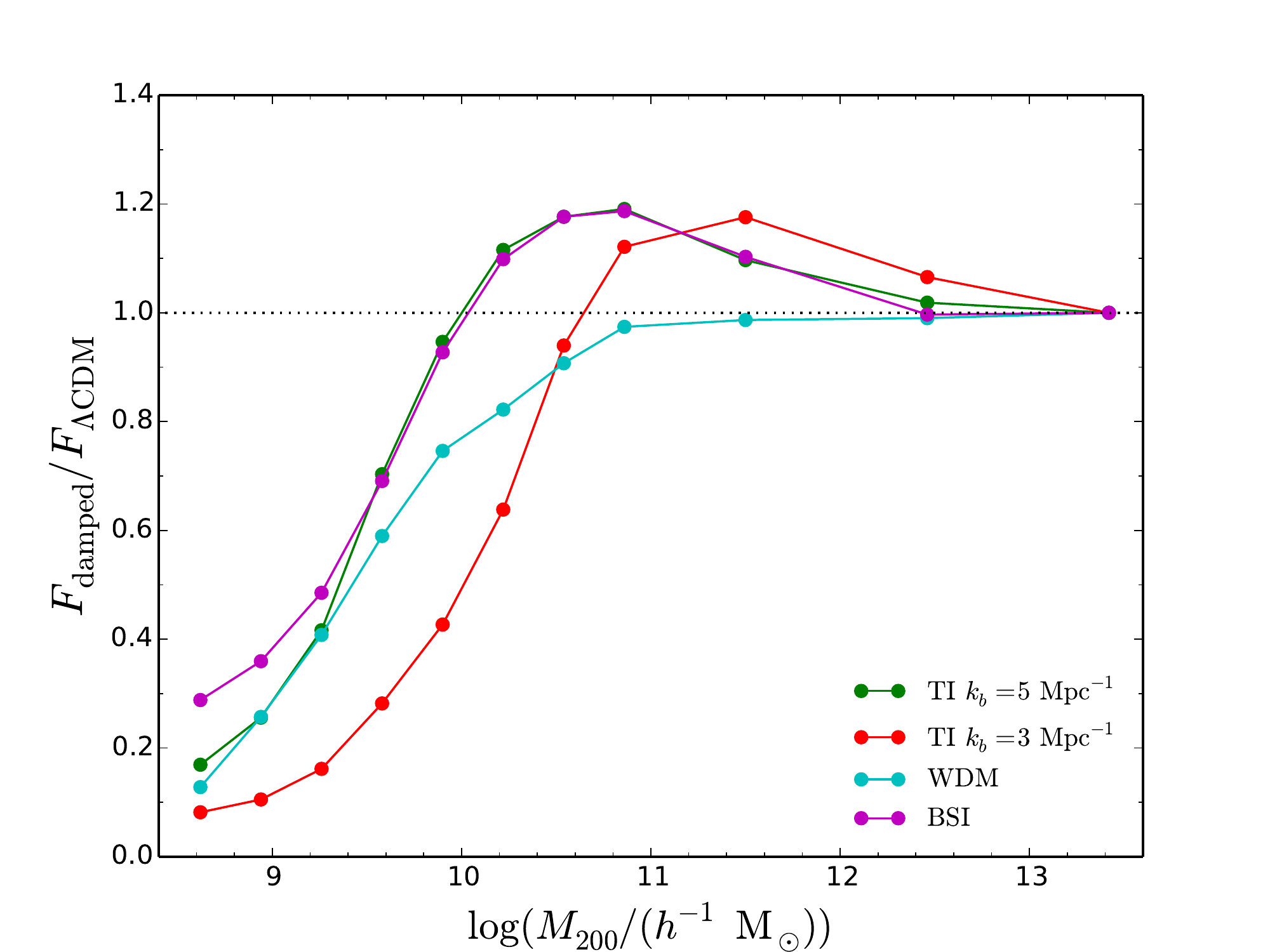}\label{fig:allhmfc}}\\
\caption{(a) Halo mass function at $z=0$ for all the models considered here (as labelled). Symbols show the results from cleaned catalogues extracted from N-body simulations and cleaned using the method explained in the text. Lines show the results obtained using the PS analytical approach with the smooth-$k$ space filter.  Note that in (a) all the halo mass functions are shifted above or below that for thermal inflation for presentation purposes. Lower panels: ratios of the halo mass function for the damped models with respect to that for $\Lambda$CDM from (b) uncleaned and (c) cleaned halo catalogues respectively. Note that in the lower panels we have reduced the number of bins in the mass range $\log(M/(h^{-1}\,\mathrm{M}_\odot)) \in[11.2,13.8]$ to suppress fluctuations due to Poisson noise.}
\label{fig:allhmf}

\end{figure}

To identify spurious haloes we adopt the method used (and extensively checked) in our previous work \cite{Leo:2018odn}. This method was proposed in \cite{Lovell:2013ola} and refines the mass criterion proposed in \cite{Wang:2007he} by excluding possible unphysical haloes by looking at the shape of the initial Lagrangian region (proto-halo) which has evolved to form a halo at late times. To decide if a halo is genuine or not, this method uses the sphericity of the proto-halo, defined as the ratio between the minor and major axes of the proto-halo region, $s \equiv c/a$. Haloes with sphericity $s< s_\mathrm{lim}$, where $s_\mathrm{lim}=0.167$, are considered to be spurious \cite{Lovell:2013ola}. We clean the halo catalogues of our simulations by considering a halo to be spurious (and then removed) if one of these conditions is satisfied \cite{Leo:2018odn}:
\begin{itemize}
\item the sphericity of the proto-halo is $s< s_\mathrm{lim}$, or
\item the halo mass is $M_\mathrm{halo}< 0.5\, M_\mathrm{lim}$.
\end{itemize}
The halo mass functions at $z=0$ extracted from the simulations are shown in Figure \ref{fig:allhmfa} for all the models considered in our analysis. The symbols show the results measured from the halo catalogues once the spurious haloes have been removed using the method described above. In the lower panels of Figure \ref{fig:allhmf} we display the measured halo mass functions (shown as ratios to the $\Lambda$CDM) for the damped models before (Figure \ref{fig:allhmfb}) and after (Figure \ref{fig:allhmfc}) the spurious structures have been removed. Comparing Figures \ref{fig:allhmfb} and \ref{fig:allhmfc} we can see that the spurious haloes affect only the low-mass end ($M_{200}<4\times 10^9\,h^{-1}\,\mathrm{M}_\odot$) of the halo mass function of the damped models. This is in line with what we found in \cite{Leo:2018odn} for nCDM models. We now discuss the cleaned results in more detail starting from the thermal inflation models and then comparing with WDM and BSI.

\emph{Thermal inflation} -- As we can see from the plots of the ratios with respect to $\Lambda$CDM (lower panels in Figure \ref{fig:allhmf}), for both thermal inflation models the halo mass function has the following behaviour: (i) it approaches the $\Lambda$CDM predictions at large halo masses, (ii) has an enhancement ($\sim 17\%$ larger than $\Lambda$CDM) at intermediate mass scales (i.e. $M_{200}\sim 3\times 10^{10}\,h^{-1}\,\mathrm{M}_\odot$ for $k_b = 5 \,\mathrm{Mpc}^{-1}$ and $M_{200}\sim 1.5\times 10^{11}\,h^{-1}\,\mathrm{M}_\odot$ for $k_b = 3 \,\mathrm{Mpc}^{-1}$) and (iii)  becomes much smaller than the $\Lambda$CDM results for lower halo masses. This behaviour follows that of the linear matter power spectrum presented in the previous section (this can be seen more clearly when using analytical approaches to calculate the halo mass function, see next subsection) and is generally expected for damped models \cite{Leo:2017wxg}.

\emph{Comparison with WDM and BSI} -- Here, we compare the halo mass function at $z=0$ for thermal inflation with $k_b= 5 \,\mathrm{Mpc}^{-1}$ with that from N-body simulations of thermal WDM and BSI inflation. As can be seen in Figure \ref{fig:allhmf}, the differences in the behaviour of the halo mass function among these three damped models follow those in the linear power spectrum (see Figure \ref{fig:linearPall}) than those in the non-linear spectrum. Indeed, at intermediate mass scales ($M_{200}\sim  10^{10}\,h^{-1}\,\mathrm{M}_\odot$) the thermal inflation and BSI halo mass functions have an enhancement with respect to $\Lambda$CDM while thermal WDM always displays halo abundances equal to or lower than standard $\Lambda$CDM. At lower masses, thermal inflation and WDM display roughly the same downturn. In the case of BSI, the halo mass function at $z=0$ is very similar to that of thermal inflation for halo masses $M_{200} > 1.5 \times 10^{9}\,h^{-1}\,\mathrm{M}_\odot$. However, at low halo masses ($M_{200}< 1.5 \times 10^{9}\,h^{-1}\,\mathrm{M}_\odot$), the halo mass function for BSI is less suppressed than that measured in the thermal inflation model. 

We note that from our N-body results (see Figure \ref{fig:allhmf}) it seems that the ratio between the BSI and $\Lambda$CDM halo mass function is close to a constant value (which is also confirmed by the analytical predictions, see next subsection) instead of decreasing further as in thermal inflation.  However, we note that our simulations cannot resolve accurately mass scales $M_{200}<5\times 10^8\,h^{-1}\,\mathrm{M}_\odot$, so higher resolution simulations would be needed to confirm the existence of this plateau at small halo masses in the BSI model with respect to $\Lambda$CDM. This different behaviour of the halo mass function at small halo masses for BSI and thermal inflation is expected from the differences at large wavenumbers in the linear power spectra of these two models. Indeed the BSI transfer function follows that of thermal inflation up to $k\sim 20 \,h\,\mathrm{Mpc}^{-1}$. However, at larger wavenumbers, the BSI transfer function does not decrease further and oscillates around a constant non-zero value, as discussed in Section 2.3 (see e.g. Figure \ref{fig:linearPall}).
\subsection{Analytical predictions}

Some aspects of the non-linear evolution of structure can be captured using semi-analytical methods. The well-known PS analytical approach is widely used to predict some  important characteristics of structure formation such as the halo mass function \cite{Press:1973iz,Sheth:1999mn,Bond:1990iw} (see also \cite{Zentner:2006vw} for a review). 

We follow the notation used in \cite{Leo:2018odn} to describe the PS approach, where the differential halo mass function is calculated as
\begin{equation}
\frac{d n}{d \log (M)} = \frac{1}{2} \,\frac{\bar{\rho}}{M} \, {f(\nu)} \frac{d \log(\nu)} {d \log(M)},
\label{eq:PShalomass}
\end{equation}
where $n$ is the halo number density, $M$ is the halo mass and $\bar{\rho}$ is the average density of the universe. $f(\nu)$ is the first-crossing distribution of  \cite{Bond:1990iw}. Assuming an ellipsoidal collapse model \cite{Sheth:1999mn}, $f(\nu)$ is well approximated by
\begin{equation}
f(\nu) = A \, \sqrt{\frac{2q\nu}{\pi}} \left( 1+ (q\nu)^{-p} \right) e^{-q\nu/2},
\end{equation}
with $A=0.3222$, $p=0.3$ and\footnote{We note that although $q=1$ is expected from a standard ellipsoidal collapse, the authors in \cite{Sheth:1999mn} found that the number of the haloes with masses $M> 10^{13}\,\mathrm{M}_\odot/h$ in $\Lambda$CDM is underpredicted, so they calibrated the value to $q=0.707$ to match N-body simulation results. Here we will maintain the standard parametrisation, $q=1$, for the following two reasons. First, when using a sharp-$k$ filter it was shown in \cite{Schneider:2013ria,Schneider:2014rda} that  $q=1$ gives a better match with simulations. We note also that the smooth-$k$ space filter was calibrated using $q=1$ \cite{Leo:2018odn}. Second, the volume of our simulations is too small to contain a statistically robust sample of such massive haloes.} $q=1$. In the above formula, $\nu$ is defined to be
\begin{equation}
\nu = \frac{\delta^2_{c,0}}{\sigma^2(R) D^2(z)},
\end{equation}
where $\delta_{c,0} = 1.686$ and $D(z)$ is the linear growth factor normalised such that $D(z=0) =1$. The variance of the density perturbations, $\sigma^2(R)$, on a given scale $R$ is
\begin{equation}
\sigma^2(R) = \int  \frac{d^3\mathbf{k}}{(2\pi)^3} P(k) \tilde{W}^2(k|R),
\label{eq:variance}
\end{equation}
where $P(k)$ is the linear matter power spectrum at $z=0$ and $\tilde{W}(k|R)$ is a filter function in Fourier space. We note that to calculate the variance in eq. (\ref{eq:variance}) we need only the linear power spectrum. This is because the halo mass function is more sensitive to the linear power spectrum than to the non-linear one. The filter function is not fixed a priori, so it could be chosen to suit the particular cosmological model and power spectrum. In simulations of standard $\Lambda$CDM, the filter function is generally chosen to be a top-hat function in real space,
\begin{equation}
W_\mathrm{Top-Hat}(x|R) = \begin{cases}
\frac{3}{4\pi R^3}\quad& \,\mathrm{if}\quad x \leq R\\
\,\,\,\,0 \quad& \,\mathrm{if}\quad x > R\\
\end{cases}, 
\end{equation}
which in Fourier space becomes,
\begin{equation}
\tilde{W}_\mathrm{Top-Hat}(k|R) = \frac{3\left(\sin(kR)-k R \cos(kR)\right)}{(kR)^3}.
\end{equation}
Other choices made in the literature include the Gaussian function and the sharp-$k$ filter (see e.g. \cite{Bond:1990iw,Zentner:2006vw}). The filter function is, in general, associated with a volume, $V_W$. In the case of a real space top-hat function, the filter in real space describes a sphere of radius $R$, so the filter volume is $V_W  = 4\pi R^3/3$, leading to a straightforward relation between the scale radius $R$ and the enclosed mass $M(R)$ of the virialised object, $M(R) = 4\pi\bar{\rho} R^3/3$. However, for other filters there is either no fixed radius in real space (e.g. for the case of a Gaussian filter) or there is a divergent integral (for a sharp-$k$ space filter)  \cite{Maggiore:2009rv}, so the mass-radius relation is usually calibrated using N-body simulations \cite{Bond:1990iw}. 

However, as shown in \cite{Schneider:2013ria,Schneider:2014rda,2013MNRAS.428.1774B}, the top-hat real space filter predicts an excess of low-mass haloes when applied to models with a cut-off in the power spectrum at small scales. Indeed, it was found e.g. in \cite{Schneider:2013ria} (but see also \cite{Leo:2018odn}) that when using a top-hat real space filter the differential halo mass function, Eq.~(\ref{eq:PShalomass}), goes as $R^{-1}$ for small radii, irrespective of the linear power spectrum considered. However, as is well known, the halo mass function for damped models takes negligible values at small halo masses (i.e. small radii). To solve this issue, \cite{Schneider:2013ria,Schneider:2014rda,2013MNRAS.428.1774B} proposed using a sharp-$k$ space filter instead.

\begin{figure}[t]
\advance\leftskip-.4cm
\advance\rightskip-3cm
\subfigure[][TI with $k_b=5\,\mathrm{Mpc}^{-1}$]
{\includegraphics[width=.55\textwidth]{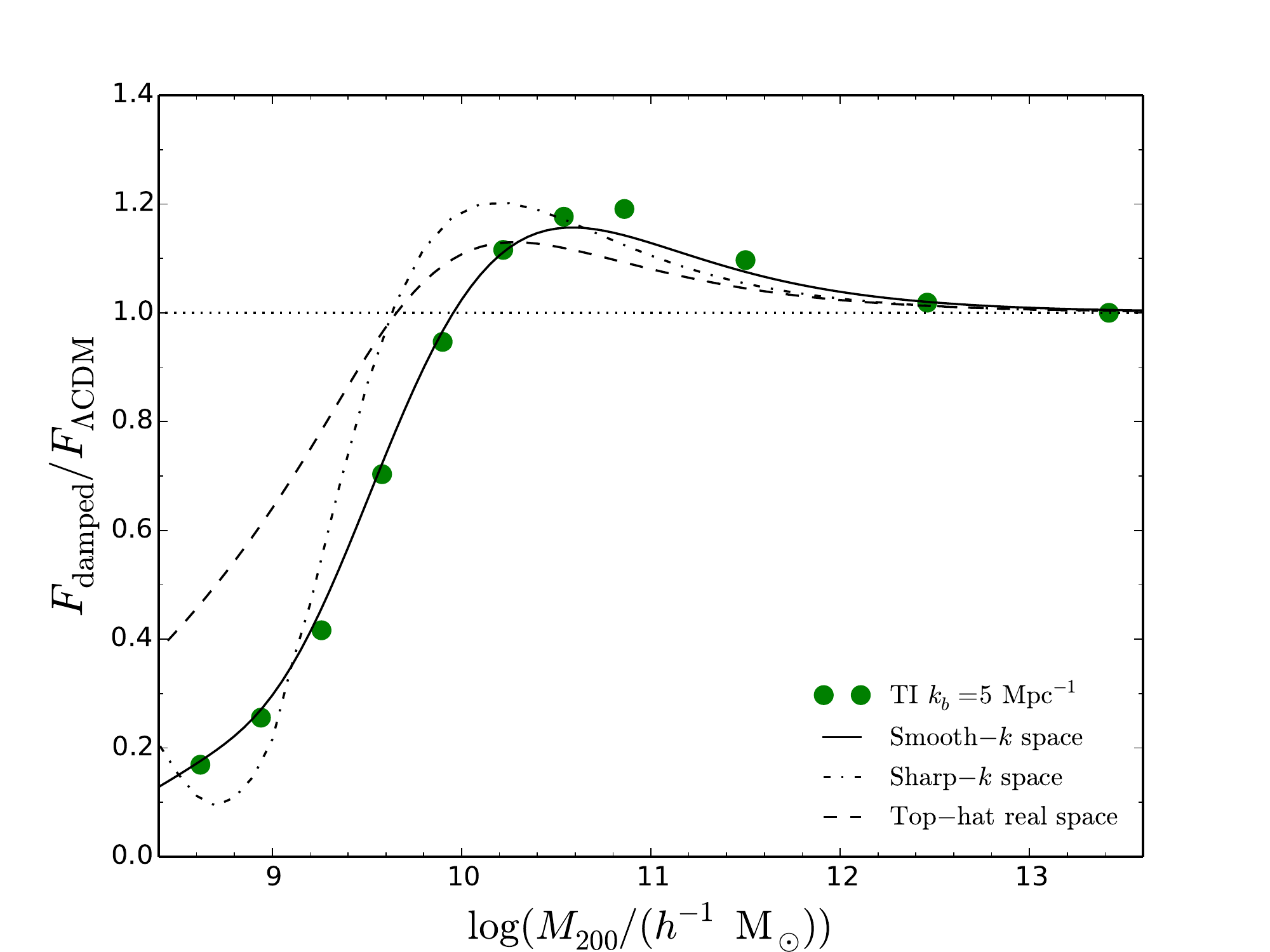}\label{fig:PShmfa}}\hspace{-1.5\baselineskip}
\subfigure[][TI with $k_b=3\,\mathrm{Mpc}^{-1}$]
{\includegraphics[width=.55\textwidth]{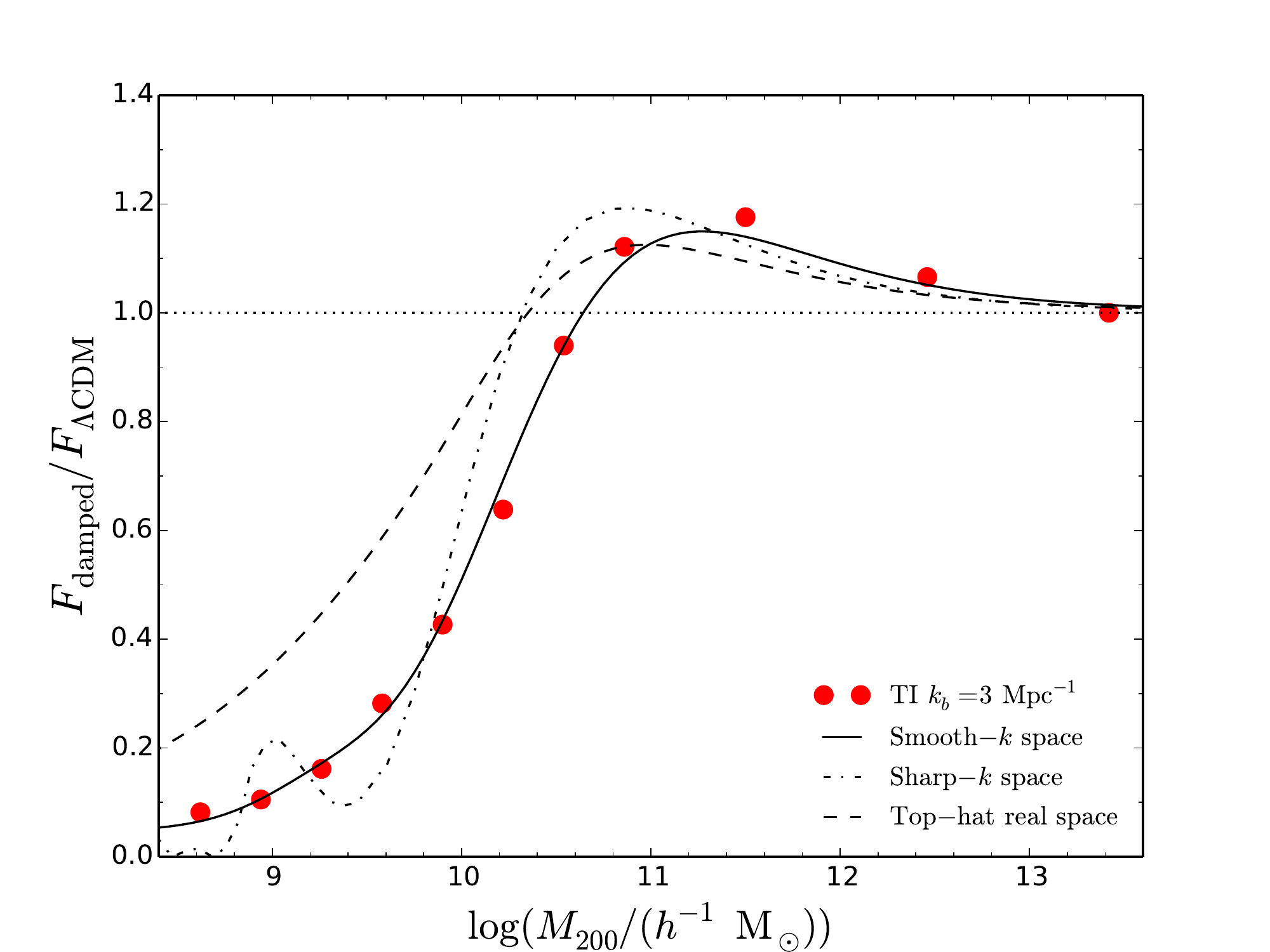}\label{fig:PShmfb}}\\
\subfigure[][WDM]
{\includegraphics[width=.55\textwidth]{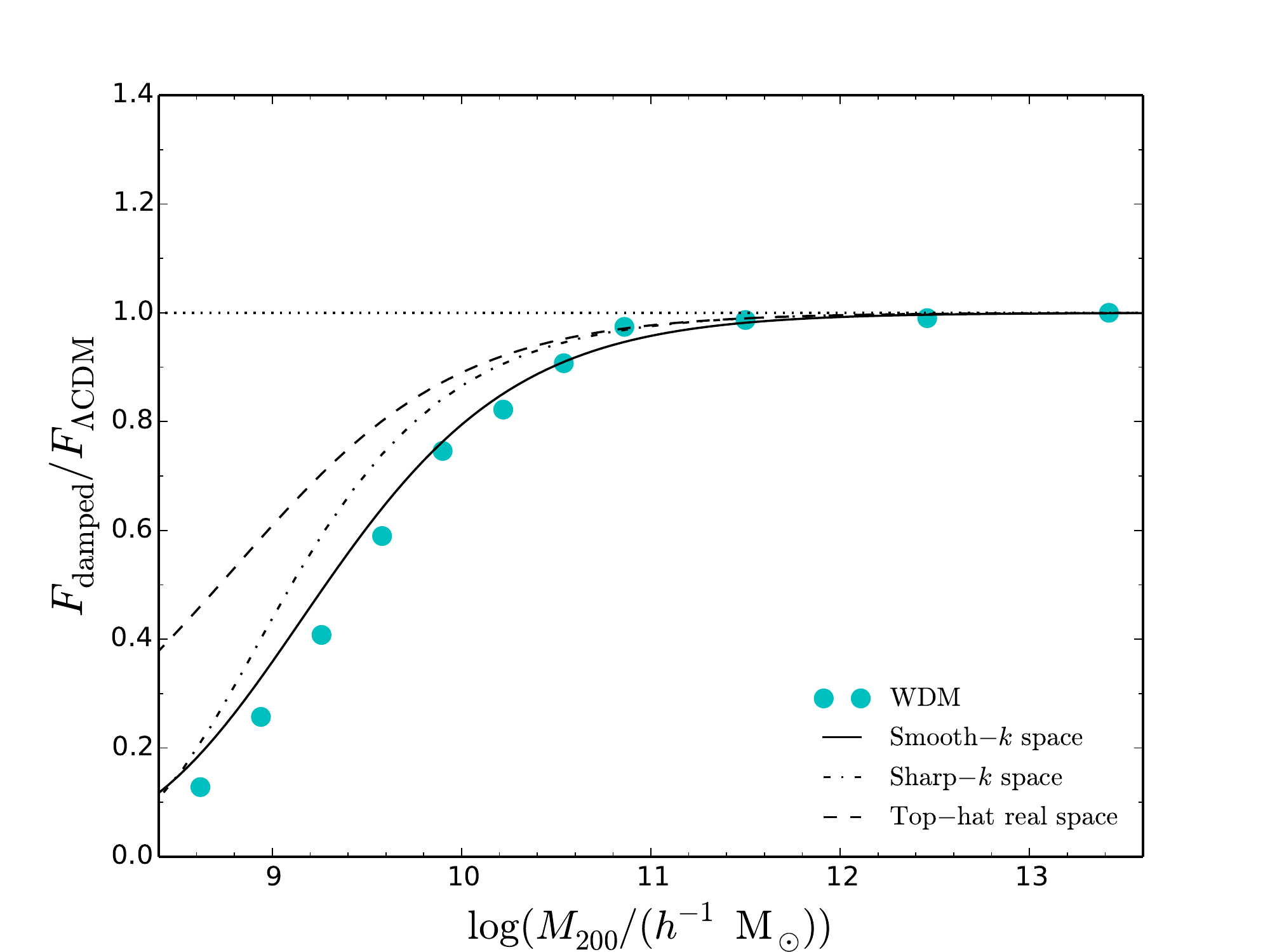}\label{fig:PShmfc}}\hspace{-1.5\baselineskip}
\subfigure[][BSI]
{\includegraphics[width=.55\textwidth]{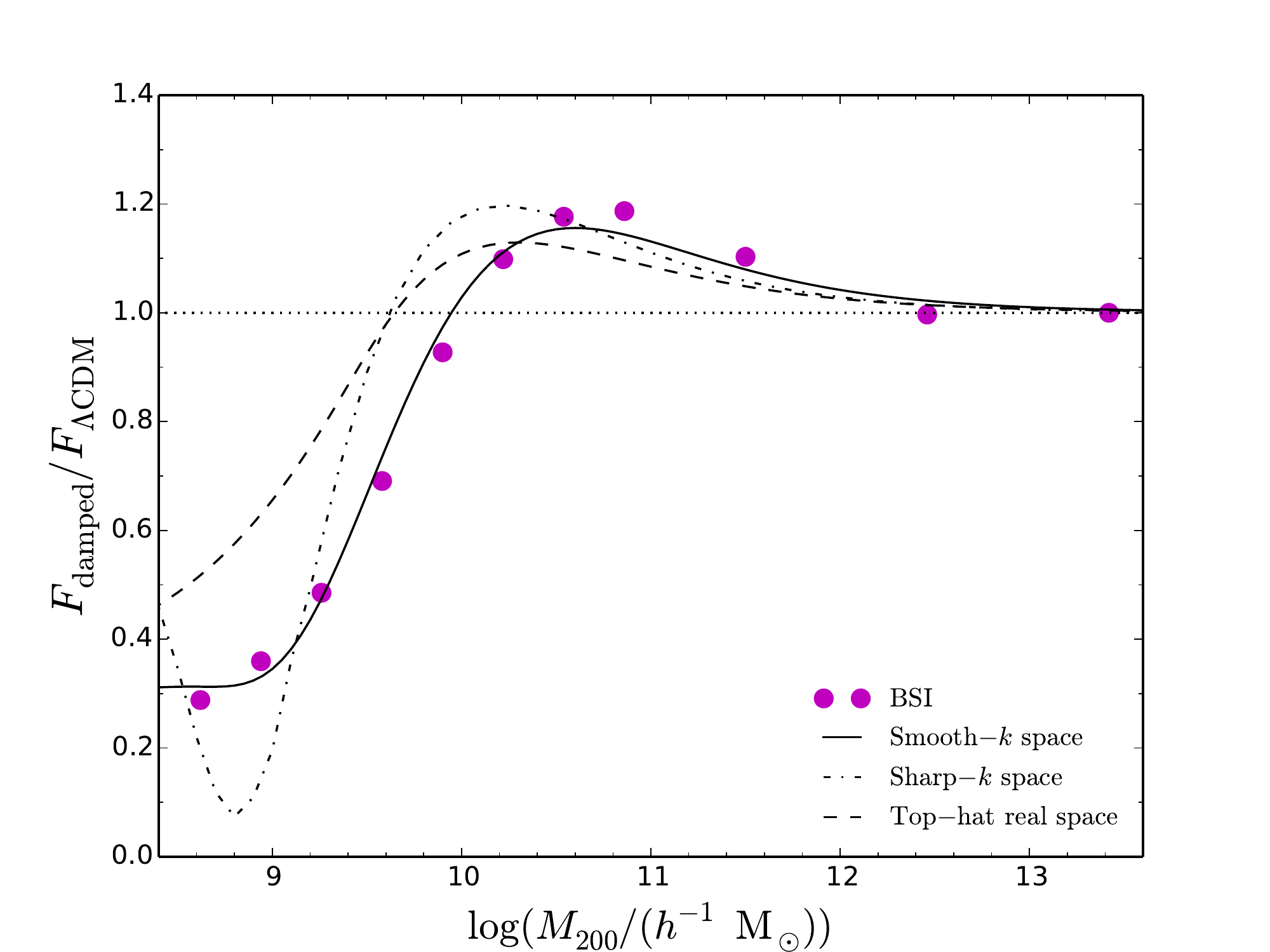}\label{fig:PShmfd}}\\
\caption{Ratios of the halo mass function at $z=0$ for the damped models with respect to that for $\Lambda$CDM. Symbols show the results from cleaned halo catalogues, while black lines show the results for the analytical PS approach with: smooth-$k$ space (solid), sharp-$k$ space (dashed-dotted) and top-hat real space (dashed) filter respectively.}
\label{fig:PShmf}

\end{figure}

In a previous work \cite{Leo:2018odn}, we have observed that, although the PS approach with a sharp-$k$ space filter reproduces very well the N-body results for thermal WDM, it does not predict correctly the halo mass function in models with a sharper truncation in the initial $P(k)$. We therefore proposed the following new filter function, dubbed the smooth-$k$ space filter, 
\begin{equation}
\tilde{W}_{\mathrm{smooth}-k}(k|R) = \left(1 + \left(kR\right)^{\hat{\beta}}\right)^{-1},
\label{eq:fittingformula2}
\end{equation}
with a mass-radius relation of the form $M(R) = \frac{4\pi}{3}\bar{\rho}(\hat{c}R)^3$, where $\{\hat{\beta},\hat{c}\}$ are free parameters. In \cite{Leo:2018odn} we found that $\{\hat{\beta}=4.8,\hat{c}=3.30\}$ give the best match with N-body simulations, so these two values will be used here as well.
This new filter overcomes the shortcomings of the sharp-$k$ space filter and gives improved agreement with N-body simulations.

We test here which of the above three filters (top-hat real space, sharp-$k$ space and smooth-$k$ space) gives the best match to the halo abundances extracted from N-body simulations. In Figure \ref{fig:PShmf}, we display as lines the results using the three filters (symbols show the results from N-body simulations). From all the panels in this figure we can see that the smooth-$k$ space filter gives better matches than the other two filters for all the models considered here (in particular the top-hat real space filter predicts an excess in the halo abundance at small halo masses of a factor of $\sim 4$ larger than that actually measured from simulations). Moreover, the smooth-$k$ space filter predicts reasonably well the position of the enhancement in the halo mass function of both thermal inflation models and BSI (with deviations $<5\%$ in the ratios). In the case of BSI, we can also see that the ratio with respect to $\Lambda$CDM predicted by our filter reaches a constant non-zero value at small halo masses. However, as pointed out in the previous subsection, our simulations cannot resolve properly these mass scales, so high-resolution simulations are needed to confirm such behaviour. We can extend the conclusions found in \cite{Leo:2018odn} by noting that the smooth-$k$ space filter is still a good choice when considering models of thermal inflation or BSI (at least for mass scales well resolved by our simulations). Here we have only considered two thermal inflation models. However, at least for all thermal inflation models with $k_b\geq 3\,\mathrm{Mpc}^{-1}$, our filter is expected to give good predictions. This is because such models are equally or less damped (more similar to $\Lambda$CDM) at the wavenumbers probed by our analysis than those considered here, so the halo abundances are less reduced at the scales relevant for structure formation.
\section{Halo statistics at high redshifts}

\begin{figure}[t]
\advance\leftskip-.4cm
\advance\rightskip-3cm
\subfigure[][Uncleaned catalogues]
{\includegraphics[width=.55\textwidth]{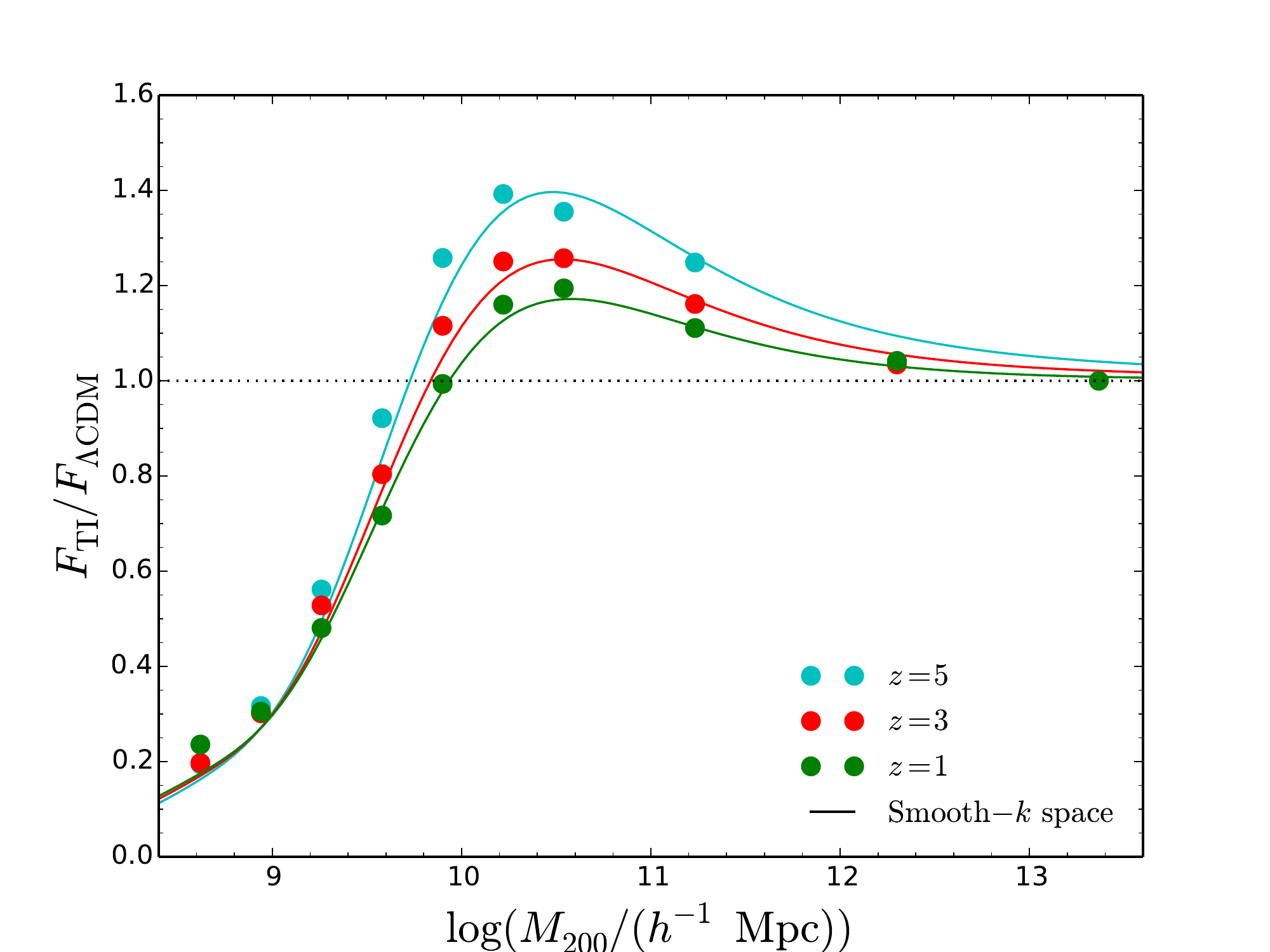}\label{fig:hmfhza}}\hspace{-1.5\baselineskip}
\subfigure[][Cleaned catalogues]
{\includegraphics[width=.55\textwidth]{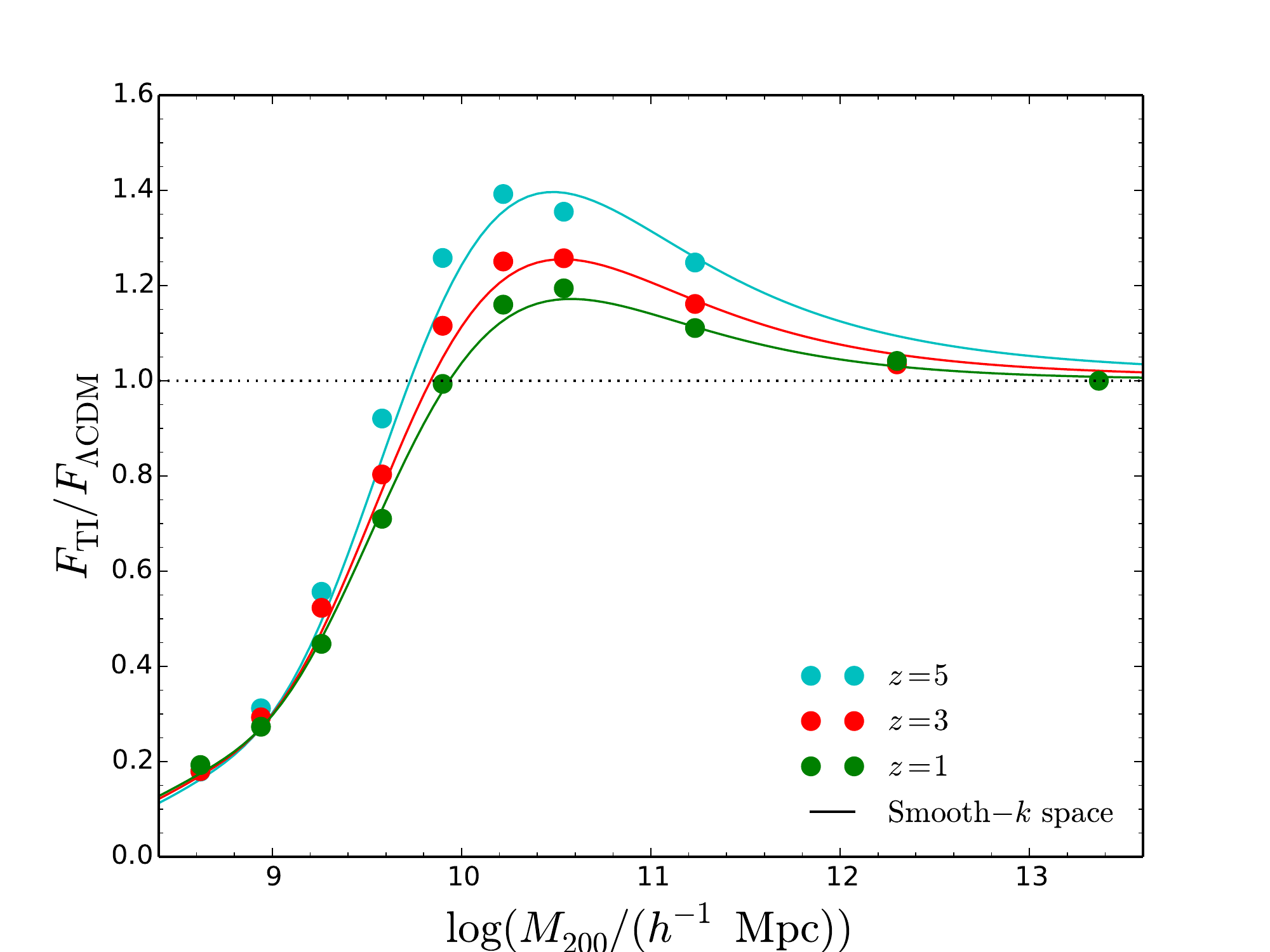}\label{fig:hmfhzb}}\\[-2. ex]
\caption{Ratios of the halo mass function for the thermal inflation with $k_b=5\,\mathrm{Mpc}^{-1}$ with respect to that for $\Lambda$CDM at different redshifts as labelled. Symbols show the results from (a) uncleaned and (b) cleaned halo catalogues. Lines show the results for the analytical PS approach with the  smooth-$k$ space filter.}
\label{fig:hmfhz}

\end{figure}

In the previous section we presented the halo abundances at $z=0$. However, in \cite{Hong:2017knn} the authors show that the magnitude of the peak in the halo mass function for thermal inflation models increases at high redshifts. In this Section we investigate this aspect by measuring the halo mass function from our N-body outputs at redshifts $z=5,3,1$ and compare them with the analytical predictions from the PS approach with the smooth-$k$ space filter. 

As in the case of $z=0$, halo catalogues from simulations at higher $z$  need to be cleaned from spurious haloes. However, our method of cleaning (considered above) uses the value in \cite{Lovell:2013ola}, $s_\mathrm{lim}=0.167$, for the sphericity limit, which is valid only for halo catalogues at $z=0$. For higher redshifts, one needs to find the appropriate sphericity limit (see e.g. the discussion in \cite{Bose:2015mga}). Since the purpose of this Section is to understand the evolution of the enhanced peak with $z$, we will not undergo a thorough study on finding the sphericity limit at $z>0$. Instead, we use a conservative estimation of the spurious haloes at high redshifts based on the haloes removed in catalogues at $z=0$. This method is illustrated below. 

For a given halo catalogue at $\tilde{z}\neq 0$, 
we first remove all the haloes with masses $M_\mathrm{halo}< 0.5\, M_\mathrm{lim}$. This is justified because we found in \cite{Leo:2017zff} that the upturn in the halo mass function due to the spurious haloes appears at roughly the same masses irrespective of the redshift. To remove further spurious haloes among those remaining  after this first cleaning step, we proceed as follows. We identify all the simulation particles belonging to spurious haloes at redshift $z=0$ (they are identified using the cleaning process explained in Section 5). We call these particles ``spurious'' particles, just because they  are in spurious haloes at $z=0$. We locate the position of the spurious particles at $\tilde{z}$ and find if these particles are bound in haloes at this redshift. If an halo at $\tilde{z}$ contains more than $70\%$ of these spurious particles, we consider this halo as spurious and we remove it from the catalogue. 

In Figure \ref{fig:hmfhz} we show the halo mass function extracted from our simulations of thermal inflation with $k_b = 5\,\mathrm{Mpc}^{-1}$ at high redshifts (the uncleaned catalogue results are shown in Figure \ref{fig:hmfhza}, while the cleaned ones are in Figure \ref{fig:hmfhzb}). From this figure we see that although the position of the enhancement in the ratio with respect to the $\Lambda$CDM remains almost invariant with the redshift, its magnitude increases with $z$. Indeed, at $z=1$ $F_\mathrm{TI}/F_{\Lambda\mathrm{CDM}}\simeq 1.19$ for masses $M_{200}\simeq 3\times 10^{10}\,h^{-1}\,\mathrm{M}_\odot$, while it increases to $\simeq 1.25$ at $z=3$ and reaches $\simeq 1.40$ at $z=5$ and for the same mass bin. We note that the differences in the halo mass function near the peak between $z=1$ and $z=0$ (see Figure \ref{fig:PShmfa} for the results at $z=0$) are instead smaller, with an enhancement in the latter of around  $F_\mathrm{TI}/F_{\Lambda\mathrm{CDM}}\simeq 1.17$ at $M_{200}\sim 3\times 10^{10}\,h^{-1}\,\mathrm{M}_\odot$ (to be compared with $F_\mathrm{TI}/F_{\Lambda\mathrm{CDM}}\simeq 1.19$ at $z=1$).  In Figure \ref{fig:hmfhz}, we show also the predictions from the PS approach with the smooth-$k$ space filter. Although the smooth-$k$ space filter parameters were calibrated using halo statistics at $z=0$ \cite{Leo:2018odn}, we can see that this filter gives reasonably accurate  predictions of the halo mass function at high redshifts. 

We found similar results in the behaviour of the enhancement at high redshifts in the thermal inflation with $k_b=3\,\mathrm{Mpc}^{-1}$ and the BSI model, while the WDM halo mass function at high redshift is always below that of $\Lambda$CDM at low masses. These results are similar to those found in Section 5 for the halo statistics at $z=0$, so we omit them here for the sake of brevity.

\section{Conclusions}
Several models which display damped matter fluctuations have been proposed to ameliorate the (possible) small-scale difficulties of the standard $\Lambda$CDM paradigm. Some of these rely on non-standard dark matter properties (such as thermal velocities or interactions), while others involve modifications to the inflationary period (such as broken scale invariance during inflation or multiple inflationary eras). 
From the point of view of structure formation, the common characteristic of all these models is the presence of a characteristic wavenumber scale above which matter fluctuations are damped. The position of this scale and the damping power depend on the particular model. 

Here we have studied, for the first time, how structures grow in the thermal inflation models described in \cite{Hong:2015oqa} by using N-body simulations. These models are characterised by a matter power spectrum which is damped on small scales but with a peak in the power (compared to $\Lambda$CDM) located at a wavenumber just below the damping scale. The N-body simulations used here were performed in a cubic box of length $L=25\,h^{-1}\,\mathrm{Mpc}$ using $512^3$ simulation particles, the simulation particle mass is around $10^7\,h^{-1}\,\mathrm{M}_\odot$. We have investigated two thermal inflation models with $k_b = 5\,\mathrm{Mpc}^{-1}$ and $k_b = 3\,\mathrm{Mpc}^{-1}$ respectively. The N-body results from such models have been compared with those from thermal WDM, BSI and standard $\Lambda$CDM model. The thermal WDM model has been chosen with the same half-mode wavenumber as in thermal inflation with $k_b = 5\,\mathrm{Mpc}^{-1}$. In the case of the BSI inflation,  we have constructed the model to have an enhanced peak at the same position and of the same magnitude as that in thermal inflation with $k_b = 5\,\mathrm{Mpc}^{-1}$. However, at large wavenumbers the linear $P(k)$ for BSI is less suppressed than that from thermal inflation. 

For each model, we have measured the non-linear power spectrum and the halo mass function from the N-body simulation outputs at different redshifts. We summarise below the main results. Regarding the nonlinear power spectrum, our findings can be summarised as follows.
\begin{itemize}
\item The peak in the linear power spectrum for both thermal inflation models persists in the nonlinear regime, but is shifted to higher wavenumbers and has a reduced amplitude at progressively lower redshifts.

\item Although the thermal inflation linear $P(k)$ for $k>3\,k_b$ is suppressed by a factor $\sim 1/50$ with respect to  $\Lambda$CDM, due to the peak shift, the non-linear power spectrum at low redshifts is enhanced at these wavenumbers with respect to that in $\Lambda$CDM. At $z=0$, the ratio w.r.t. the $\Lambda$CDM is of the order of unity for all the wavenumbers $k<k_\mathrm{Ny}$, where $k_\mathrm{Ny}$ is the Nyquist frequency of our simulations. This is true for both thermal inflation models considered here.

\item The WDM non-linear power spectrum behaves differently than that from thermal inflation. Indeed, we find that although the transfer of power enhances the power at small scales, the non-linear power spectra at low $z$ always have less power than the standard $\Lambda$CDM $P(k)$.

\item The BSI non-linear power spectrum shows  similar behaviour to that in thermal inflation. Indeed, the peak in BSI is shifted to higher wavenumbers and has a reduced amplitude at low redhsifts as in the case of thermal inflation with $k_b = 5\,\mathrm{Mpc}^{-1}$. In the linear regime, $P_\mathrm{TI}/P_\mathrm{BSI}\sim0.09$ at $k\gg k_b$. However, the non-liner evolution (almost completely) washes out the differences at large wavenumbers present in the two initial power spectra, so that at low redshifts the two non-linear power spectra are very similar to each other, e.g. $P_\mathrm{TI}/P_\mathrm{BSI}\sim0.98$ at $k\sim k_\mathrm{Ny}$.
\end{itemize}
Regarding the halo mass function, we find the following results.
\begin{itemize}
\item In general, the differences in the halo mass function between the different models studied here follow those in the linear matter power spectra (rather than those in the non-linear power spectrum). Indeed, in both the thermal inflation models we find that the halo mass function has an enhanced peak before dropping to negligible values at small masses (as in the case of the linear power spectrum). The BSI and thermal inflation with $k_b = 5\,\mathrm{Mpc}^{-1}$ halo mass functions are roughly the same for masses $M_{200}>1.5 \times 10^{9}\,h^{-1}\,\mathrm{M}_\odot$. At lower masses, the BSI decreases slower than thermal inflation and eventually reaches a plateau. In the case  of WDM, the halo mass function presents no enhancements and is always equal or less than that from the standard $\Lambda$CDM model. 
\item The enhancement in the halo mass function is around $M_{200}\sim 3\times 10^{10}\,h^{-1}\,\mathrm{M}_\odot$ for thermal inflation with $k_b = 5\,\mathrm{Mpc}^{-1}$ and BSI models, while it appears at higher masses, $M_{200}\sim 1.5\times 10^{11}\,h^{-1}\,\mathrm{M}_\odot$, for $k_b = 3 \,\mathrm{Mpc}^{-1}$. The enhancement in all these models with respect to the $\Lambda$CDM is around $\sim 20\%$ at $z=0$.
\item At higher redshifts and in the case of thermal inflation, we find that the halo mass function has an enhancement at the same mass as $z=0$. However, the magnitude of this enhancement with respect to $\Lambda$CDM increases from $\sim 20\%$ at $z=0$ to $\sim 40\%$ at $z=5$.
\item We have used these numerical results to test  the predictions from the PS approach with three filters: top-hat real space, sharp-$k$ space and smooth-$k$ space. From this analysis, we find that the predictions from a smooth-$k$ space filter agree with the simulation results over the widest range of halo masses. This is true for all the models and at all the redshifts considered here.
\end{itemize} 
We note that here we have considered the simplest WDM model, the so-called thermal WDM. However, as mentioned in the introduction, this model does not exhaust all the possibilities inside the nCDM scenarios. Other well-motivated models, such as sterile neutrinos \cite{Dodelson:1993je,Dolgov:2000ew,Asaka:2006nq,Enqvist:1990ek,Shi:1998km,Abazajian:2001nj,Kusenko:2006rh,Petraki:2007gq,Merle:2015oja,Konig:2016dzg}, axion-like particles \cite{Marsh:2015xka,Veltmaat:2016rxo,Veltmaat:2018dfz} or models suggested by effective theory of structure formation (ETHOS) \cite{Cyr-Racine:2015ihg,Vogelsberger:2015gpr}, can produce different effects on structure formation than those found in the case of thermal WDM (see e.g. \cite{Murgia:2017lwo,Veltmaat:2016rxo,Veltmaat:2018dfz,Lovell:2017eec,Lovell:2015psz}). We note also that our simulations do not take into account the effects of baryon physics that can modify the matter distribution and the properties of haloes measured from DM-only simulations, see e.g. \cite{vanDaalen:2013ita,Sawala:2015cdf,Hellwing:2016ucy,2017MNRAS.468.4285L,Artale:2016rit,Vogelsberger:2014dza,Springel:2017tpz}.

As in nCDM models, the thermal inflation model predicts fewer low-mass haloes and sub-haloes than standard $\Lambda$CDM, but without involving any modification to the dark matter sector. However, differently from nCDM, the model presents an enhancement in the power spectrum at wavenumbers just shorter than the damping scale. This is a unique feature in the power spectrum which could leave an imprint on the large-scale structure of the Universe which can be used to distinguish this model from nCDM scenarios. For example, in \cite{Hong:2017knn} the authors have shown how future 21-cm observations could be able to distinguish between thermal inflation and WDM because of the different shapes of their power spectra. The recent results from the EDGES experiment \cite{2018Natur.555...67B}, which claim a detection of a 21-cm absorption line at $z\sim 20$, could indicate that star formation was happening at high redshifts (an early onset of the so-called cosmic dawn). An early cosmic dawn can be used to constrain the suppression of the matter power spectrum in damped models (as was done e.g. for nCDM models in \cite{Schneider:2018xba,Lidz:2018fqo}) and these constraints can be applied to thermal inflation power spectra. Moreover, Lyman-$\alpha$ observations  could be used  to constrain the values of $k_b$ as was done for the mass of thermal WDM candidates in \cite{Viel:2013apy,Irsic:2017ixq}. Regarding halo abundances, the halo mass function in thermal inflation is characterised by an enhanced abundance at halo masses just above the suppression mass scale, so in principle, these models can be distinguished from nCDM using halo statistics. Indeed, galaxy probes (see e.g.  \cite{Gardner:2006ky,2016SPIE.9908E..1ZD}) at high redshift could be able to give some information on the massive objects near the peak (which is enhanced in magnitude for $z>0$ as we have found here), while future strong lensing observations (see e.g. \cite{Li:2015xpc}) will be able to constrain the number of low-mass DM sub-haloes. Results from these future observations could shed some light on the shape of the underlying matter power spectrum and could indirectly constrain the nature of the processes that produce a damping in the matter fluctuation from very early epochs in the history of the Universe.

\acknowledgments

ML and SP are supported by the European Research Council under ERC Grant ``NuMass'' (FP7- IDEAS-ERC ERC-CG 617143). ML and BL are supported by an European Research Council Starting Grant (ERC-StG-716532-PUNCA). CMB and BL acknowledge the support of the UK STFC Consolidated Grants (ST/P000541/1 and ST/L00075X/1) and Durham University. SP acknowledges partial support from the Wolfson Foundation and the Royal Society and also thanks SISSA  and IFT UAM-CSIC for support and hospitality during part of this work. SP, CMB and BL are also supported in part by the European Union's Horizon 2020 research and innovation program under the Marie Sk\l{}odowska-Curie grant agreements No. 690575 (RISE InvisiblesPlus) and 674896 (ITN Elusives). This work used the DiRAC Data Centric system at Durham University, operated by the Institute for Computational Cosmology on behalf of the STFC DiRAC HPC Facility (\href{www.dirac.ac.uk}{www.dirac.ac.uk}). This equipment was funded by BIS National E-infrastructure capital grant ST/K00042X/1, STFC capital grants ST/H008519/1 and ST/K00087X/1, STFC DiRAC Operations grant  ST/K003267/1 and Durham University. DiRAC is part of the National E-Infrastructure.

\appendix
\section{Numerical convergence}
\label{sec:numconv}

\begin{figure}[h!]
\centering
{\includegraphics[width=.7\textwidth]{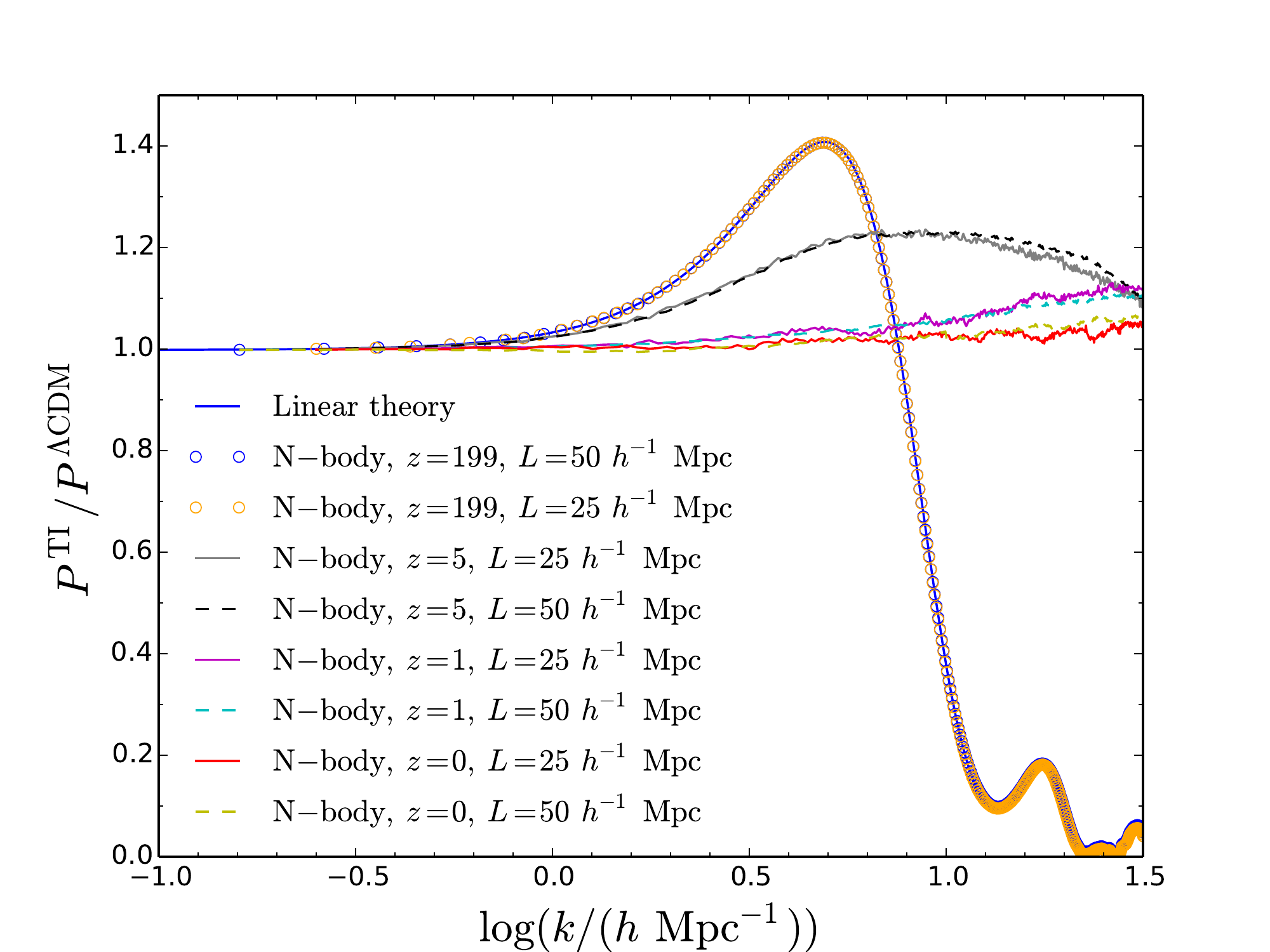}}
\caption{Ratios of the matter power spectrum measured from simulations of thermal inflation with $k_b=3\,\mathrm{Mpc}^{-1}$ with respect to that from standard $\Lambda$CDM at redshifts $z=199,5,1,0$. Solid curves represent simulations with $L=25 \,h^{-1}\,\mathrm{Mpc}$, while dashed curves show simulations with $L=50 \,h^{-1}\,\mathrm{Mpc}$. The solid blue line shows the linear theory prediction and is mostly obscured by the orange symbols. Blue symbols show results from ICs at $z=199$ for simulations with $L=50 \,h^{-1}\,\mathrm{Mpc}$, while orange symbols are from ICs of simulations with $L=25 \,h^{-1}\,\mathrm{Mpc}$.}
\label{fig:resolPk}
\end{figure}

\begin{figure}
\centering
{\includegraphics[width=.7\textwidth]{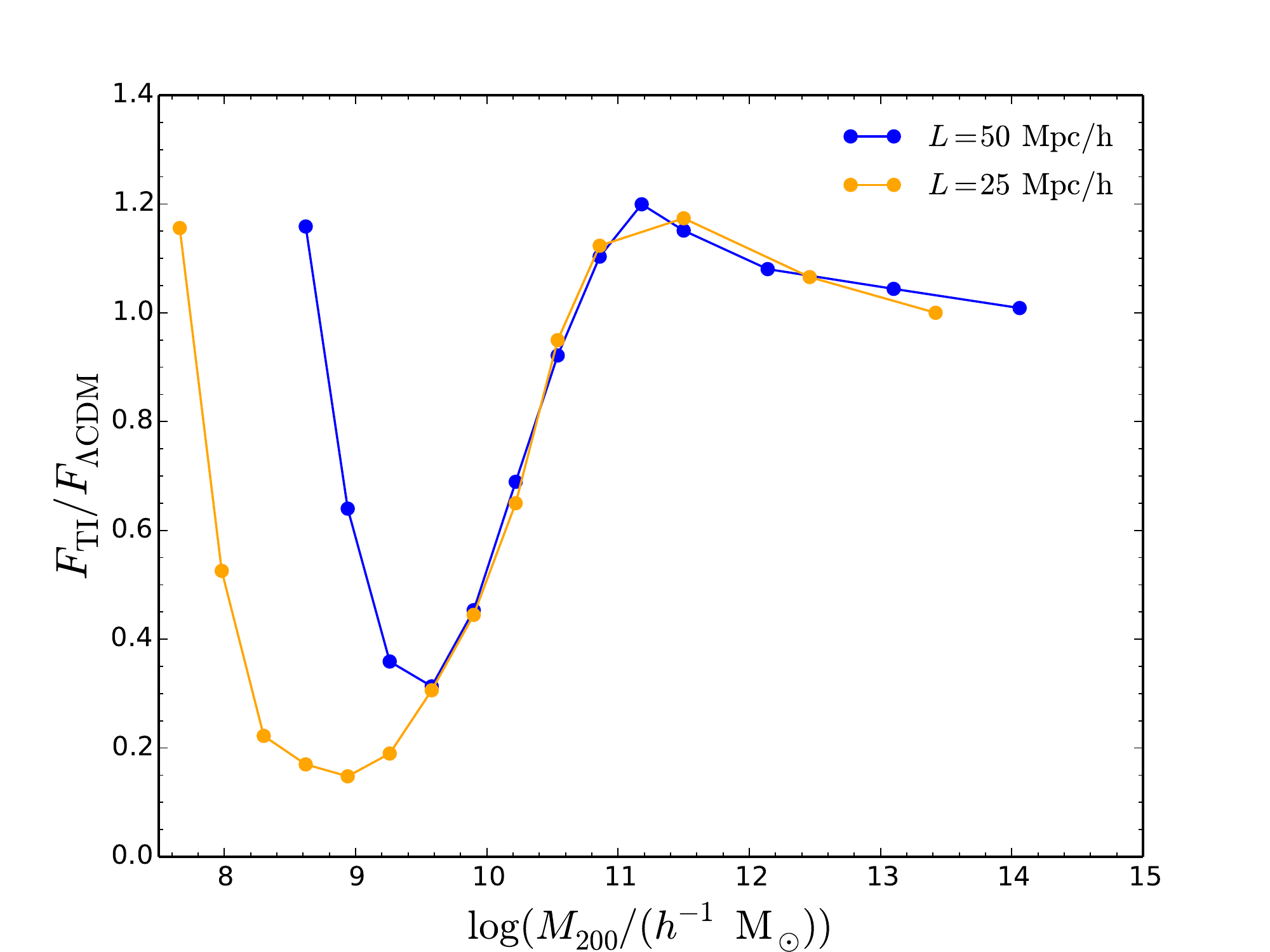}}
\caption{Ratio of the halo mass function measured from simulations of thermal inflation with $k_b=3\,\mathrm{Mpc}^{-1}$ with respect to that from standard $\Lambda$CDM at redshift $z=0$. Blue symbols  show results from simulations with $L=50 \,h^{-1}\,\mathrm{Mpc}$, while orange symbols show results from the smaller box, $L=25 \,h^{-1}\,\mathrm{Mpc}$.}
\label{fig:resolHMF}
\end{figure}

In this appendix, we study the accuracy of simulations with box length and particle number $\{L=25 \,h^{-1}\,\mathrm{Mpc}, N=512^3\}$. We have run another set of  simulations with parameters $\{L=50 \,h^{-1}\,\mathrm{Mpc}, N=512^3\}$, for both the standard $\Lambda$CDM and the thermal inflation with $k_b=3\,\mathrm{Mpc}^{-1}$. In Figure \ref{fig:resolPk} 
 we show the ratio of the power spectrum measured from the thermal inflation model with respect to that from standard $\Lambda$CDM at redshifts $z=5,1$ and $0$ for simulations with $L=50\,h^{-1}\,\mathrm{Mpc}$ (dashed curves) and $L=25 \,h^{-1}\,\mathrm{Mpc}$ (solid curves). As can be seen from the figure, the power spectra measured from the simulations with different box lengths are in good agreement at small wavenumbers (up to the Nyquist frequency of the larger box simulation), this is also true for other redshifts which we omit for the sake of brevity. 
 
A similar analysis can be done for the halo mass function. In Figure \ref{fig:resolHMF} we show the results at $z=0$ for this quantity measured from simulations with  $L=50\,h^{-1}\,\mathrm{Mpc}$ and $L=25 \,h^{-1}\,\mathrm{Mpc}$. As we can see, the numerical results from the two boxes converge at large halo masses, $M_{200}>10^{10}\,h^{-1}\,\mathrm{M}_\odot$. At lower masses the results from the larger box are dominated by spurious structures as expected.



\newpage
\begin{thebibliography}{99}
\bibitem{Weinberg:2013aya} D. H. Weinberg, J. S. Bullock, F. Governato, R. K. de Naray, A. H. G. Peter, \emph{Cold dark matter: controversies on small scales}, Proc. Nat. Acad. Sci. {\bf 112}, 12249-12255 (2014) [\href{https://arxiv.org/abs/1306.0913}{arXiv:1306.0913} [astro-ph.CO]].
\bibitem{2012MNRAS.421.3464P}  A. Pontzen, F. Governato, \emph{How supernova feedback turns dark matter cusps into cores}, Mon. Not. R. Astron. Soc. {\bf 421}, 3464-3471 (2012)   [\href{https://arxiv.org/abs/1106.0499}{arXiv:1106.0499} [astro-ph.CO]].
\bibitem{2013MNRAS.432.1947M} D. Martizzi, R. Teyssier, B. Moore, \emph{Cusp-core transformations induced by AGN feedback in the progenitors of cluster galaxies}, Mon. Not. R. Astron. Soc. {\bf 432}, 1947-1954 (2013)  [\href{https://arxiv.org/abs/1211.2648}{arXiv:1211.2648} [astro-ph.CO]].
\bibitem{2014ApJ...786...87B} A. M. Brooks, A. Zolotov, \emph{Why Baryons Matter: The Kinematics of Dwarf Spheroidal Satellites}, ApJ {\bf 786}, 87 (2014)  [\href{https://arxiv.org/abs/1207.2468}{arXiv:1207.2468} [astro-ph.CO]].
\bibitem{2012MNRAS.424.2715W} J. Wang, C. S. Frenk, J. F. Navarro, L. Gao, T. Sawala, \emph{The missing massive satellites of the Milky Way}, Mon. Not. R. Astron. Soc. {\bf 424},  2715-2721 (2012)   [\href{https://arxiv.org/abs/1203.4097}{arXiv:1203.4097} [astro-ph.GA]].

\bibitem{Kamionkowski:1999vp} M. Kamionkowski, A. R. Liddle, \emph{The Dearth of halo dwarf galaxies: Is there power on short scales?}, Phys. Rev. Lett.  {\bf 84}, 4525 (2000) [\href{https://arxiv.org/abs/astro-ph/9911103}{arXiv:astro-ph/9911103}].
\bibitem{White:2000sy} M. J. White, R. A. C. Croft, \emph{Suppressing linear power on dwarf galaxy halo scales}, Astrophys. J.  {\bf 539}, 497 (2000) [\href{https://arxiv.org/abs/astro-ph/0001247v2}{arXiv:astro-ph/0001247}].
\bibitem{Yokoyama:2000tz} J. Yokoyama, \emph{Inflation and the dwarf galaxy problem}, Phys. Rev. D {\bf 62}, 123509 (2000) [\href{https://arxiv.org/abs/astro-ph/0009127}{arXiv:astro-ph/0009127}].  
\bibitem{Zentner:2002xt} A. R. Zentner, J. S. Bullock, \emph{Inflation, cold dark matter, and the central density problem}, Phys. Rev. D {\bf 66}, 043003 (2002) [\href{https://arxiv.org/abs/astro-ph/0205216}{arXiv:astro-ph/0205216}].
\bibitem{Ashoorioon:2006wc}  A. Ashoorioon, A. Krause, \emph{Power Spectrum and Signatures for Cascade Inflation}. [\href{https://arxiv.org/abs/hep-th/0607001}{arXiv:hep-th/060700}].
\bibitem{Kobayashi:2010pz} T. Kobayashi, F. Takahashi, \emph{Running Spectral Index from Inflation with Modulations}, JCAP {\bf 1101}, 026 (2011) [\href{https://arxiv.org/abs/1011.3988}{arXiv:1011.3988} [astro-ph.CO]].
\bibitem{Nakama:2017ohe} T. Nakama, J. Chluba, M. Kamionkowski, \emph{Shedding light on the small-scale crisis with CMB spectral distortions}, Phys. Rev. D {\bf 95}, no. 12, 121302 (2017)  [\href{https://arxiv.org/abs/1703.10559v2}{arXiv:1703.10559} [astro-ph.CO]].
\bibitem{Hong:2015oqa} 
  S. E. Hong, H. J. Lee, Y. J. Lee, E. D. Stewart, H. Zoe,
  \emph{Effects of thermal inflation on small scale density perturbations},
  JCAP {\bf 1506}, 002 (2015)
  [\href{https://arxiv.org/abs/1503.08938}{arXiv:1503.08938} [astro-ph.CO]].
\bibitem{Starobinsky:1992ts} 
  A. A. Starobinsky,
  \emph{Spectrum of adiabatic perturbations in the universe when there are singularities in the inflation potential},
  JETP Lett.\  {\bf 55}, 489 (1992). 
\bibitem{Murgia:2017lwo} 
  R. Murgia, A. Merle, M. Viel, M. Totzauer, A. Schneider,
  \emph{"Non-cold" dark matter at small scales: a general approach},
  JCAP {\bf 1711}, 046 (2017)
  [\href{https://arxiv.org/abs/1704.07838}{arXiv:1704.07838} [astro-ph.CO]].  
\bibitem{Bode:2000gq} P. Bode, J. P. Ostriker, N. Turok, \emph{Halo formation in warm dark matter models}, Astrophys. J. {\bf 556}, 93-107 (2001)  [\href{http://arxiv.org/abs/astro-ph/0010389v3}{arXiv:astro-ph/0010389}].
\bibitem{Colin:2000dn} P. Colin, V. Avila-Reese, O. Valenzuela, \emph{Substructure and halo density profiles in a warm dark matter cosmology}, Astrophys. J. {\bf 542}, 622-630 (2000)  [\href{https://arxiv.org/abs/astro-ph/0004115}{arXiv:astro-ph/0004115}].
\bibitem{Hansen:2001zv} S. H. Hansen, J. Lesgourgues, S. Pastor, J. Silk, \emph{Constraining the window on sterile neutrinos as warm dark matter}, Mon. Not. Roy. Astron. Soc. {\bf 333}, 544-546 (2002)  [\href{https://arxiv.org/abs/astro-ph/0106108}{arXiv:astro-ph/0106108}].
\bibitem{Viel:2005qj} M.Viel, J. Lesgourgues, M. G. Haehnelt, S. Matarrese, A. Riotto, \emph{Constraining warm dark matter candidates including
                        sterile neutrinos and light gravitinos with WMAP and the
                        Lyman-alpha forest}, Phys. Rev. D {\bf 71}, 063534 (2005)  [\href{https://arxiv.org/abs/astro-ph/0501562}{arXiv:astro-ph/0501562}].
\bibitem{Dodelson:1993je} S. Dodelson, L. M. Widrow, \emph{Sterile-neutrinos as dark matter}, Phys. Rev. Lett. {\bf 72}, 17-20 (1994)  [\href{http://xxx.lanl.gov/abs/hep-ph/9303287}{arXiv:hep-ph/9303287}].
\bibitem{Dolgov:2000ew} A. D. Dolgov, S. H. Hansen, \emph{Massive sterile neutrinos as warm dark matter}, Astropart. Phys. {\bf 16}, 339-344 (2002)   [\href{https://arxiv.org/abs/hep-ph/0009083}{arXiv:hep-ph/0009083}].
\bibitem{Asaka:2006nq} T. Asaka, M. Laine, M. Shaposhnikov, \emph{Lightest sterile neutrino abundance within the nuMSM}, JHEP {\bf 01}, 091 (2007)  [Erratum: JHEP02,028(2015)] [\href{http://xxx.lanl.gov/abs/hep-ph/0612182}{arXiv:hep-ph/0612182}].
\bibitem{Enqvist:1990ek}  K. Enqvist, K. Kainulainen, J. Maalampi, \emph{Resonant neutrino transitions and nucleosynthesis}, Phys. Lett. B {\bf 249},  531-534 (1990).
\bibitem{Shi:1998km} X. Shi, G. M. Fuller, \emph{A New dark matter candidate: Nonthermal sterile
                        neutrinos}, Phys. Rev. Lett. {\bf 82}, 2832 (1999)  [\href{http://xxx.lanl.gov/abs/astro-ph/9810076}{arXiv:astro-ph/9810076}].
\bibitem{Abazajian:2001nj} K. Abazajian, G. M. Fuller, M. Patel, \emph{Sterile neutrino hot, warm, and cold dark matter}, Phys. Rev. D {\bf 64},  023501 (2001)  [\href{http://xxx.lanl.gov/abs/astro-ph/0101524}{arXiv:astro-ph/0101524}]. 
\bibitem{Kusenko:2006rh} A. Kusenko, \emph{Sterile neutrinos, dark matter, and the pulsar
                        velocities in models with a Higgs singlet}, Phys. Rev. Lett. {\bf 97}, 241301 (2006)  [\href{https://arxiv.org/abs/hep-ph/0609081}{arXiv:hep-ph/0609081}].
\bibitem{Petraki:2007gq} K. Petraki, A. Kusenko, \emph{Dark-matter sterile neutrinos in models with a gauge
                        singlet in the Higgs sector}, Phys. Rev. D {\bf 77}, 065014 (2008)  [\href{https://arxiv.org/abs/0711.4646}{arXiv:0711.4646} [hep-ph]].
\bibitem{Merle:2015oja} A. Merle, M. Totzauer, \emph{keV Sterile Neutrino Dark Matter from Singlet Scalar
                        Decays: Basic Concepts and Subtle Features}, JCAP {\bf 1506}, 011 (2015)  [\href{https://arxiv.org/abs/1502.01011}{arXiv:1502.01011} [hep-ph]].
\bibitem{Konig:2016dzg}  J. K\"{o}nig, A. Merle, M. Totzauer, \emph{keV Sterile Neutrino Dark Matter from Singlet Scalar
                        Decays: The Most General Case}, JCAP {\bf 1611}, 038 (2016)  [\href{https://arxiv.org/abs/1609.01289}{arXiv:1609.01289} [hep-ph]].
\bibitem{Boehm:2004th} C. Boehm, R. Schaeffer, \emph{Constraints on dark matter interactions from structure formation: Damping lengths}, Astron. Astrophys. {\bf 438}, 419-442 (2005)  [\href{https://arxiv.org/abs/astro-ph/0410591}{arXiv:astro-ph/0410591}]. 
\bibitem{Boehm:2014vja} C. Boehm, J. A. Schewtschenko, R. J. Wilkinson, C. M. Baugh, S. Pascoli, \emph{Using the Milky Way satellites to study interactions
                        between cold dark matter and radiation}, Mon. Not. Roy. Astron. Soc. {\bf 445},  L31-L35 (2014)  [\href{https://arxiv.org/abs/1404.7012}{arXiv:1404.7012} [astro-ph.CO]].
\bibitem{Schewtschenko:2014fca} J. A. Schewtschenko, R. J. Wilkinson, C. M. Baugh, C. Boehm, S. Pascoli, \emph{Dark matter-radiation interactions: the impact on dark matter haloes}, Mon.\ Not.\ Roy.\ Astron.\ Soc.\  {\bf 449}, 3587 (2015)

\bibitem{Spergel:1999mh} D. N. Spergel, P. J. Steinhardt, \emph{Observational evidence for selfinteracting cold dark
                        matter}, Phys. Rev. Lett. {\bf 84}, 3760-3763 (2000)  [\href{https://arxiv.org/abs/astro-ph/9909386}{arXiv:astro-ph/9909386}].
\bibitem{Marsh:2015xka} D. J. E. Marsh, \emph{Axion Cosmology}, Phys. Rept. {\bf 643}, 1-79 (2016)  [\href{https://arxiv.org/abs/1510.07633}{arXiv:1510.07633} [astro-ph.CO]].
\bibitem{Veltmaat:2016rxo} 
  J. Veltmaat, J. C. Niemeyer,
  \emph{Cosmological particle-in-cell simulations with ultralight axion dark matter},
  Phys. Rev. D {\bf 94}, no. 12, 123523 (2016)
  [\href{https://arxiv.org/abs/1608.00802}{arXiv:1608.00802} [astro-ph.CO]].
\bibitem{Veltmaat:2018dfz} 
  J. Veltmaat, J. C. Niemeyer, B. Schwabe,
  \emph{Formation and structure of ultralight bosonic dark matter halos}, (2018)
  [\href{https://arxiv.org/abs/1804.09647}{arXiv:1804.09647} [astro-ph.CO]].
\bibitem{Lyth:1995hj} 
  D. H. Lyth, E. D. Stewart,
  \emph{Cosmology with a TeV mass GUT Higgs},
  Phys. Rev. Lett.  {\bf 75}, 201 (1995)
 [\href{https://arxiv.org/abs/hep-ph/9502417}{arXiv:hep-ph/9502417}].
\bibitem{Lyth:1995ka} 
  D. H. Lyth, E. D. Stewart,
  \emph{Thermal inflation and the moduli problem},
  Phys. Rev. D {\bf 53}, 1784 (1996)
  [\href{https://arxiv.org/abs/hep-ph/9510204}{hep-ph/9510204}].
\bibitem{Banks:1993en} 
  T. Banks, D. B. Kaplan, A. E. Nelson,
  \emph{Cosmological implications of dynamical supersymmetry breaking},
  Phys. Rev. D {\bf 49}, 779 (1994)
  [\href{https://arxiv.org/abs/hep-ph/9308292}{arXiv:hep-ph/9308292}].

\bibitem{deCarlos:1993wie} 
  B. de Carlos, J. A. Casas, F.~Quevedo, E. Roulet,
  \emph{Model independent properties and cosmological implications of the dilaton and moduli sectors of 4-d strings},
  Phys. Lett. B {\bf 318}, 447 (1993)
  [\href{https://arxiv.org/abs/hep-ph/9308325}{arXiv:hep-ph/9308325}].
\bibitem{Cho:2017zkj} 
  K. Cho, S. E. Hong, E. D. Stewart, H.~Zoe,
  \emph{CMB Spectral Distortion Constraints on Thermal Inflation},
  JCAP {\bf 1708}, no. 08, 002 (2017)
 [\href{https://arxiv.org/abs/1705.02741}{arXiv:1705.02741} [astro-ph.CO]].
\bibitem{Hong:2017knn} 
  S. E. Hong, H. Zoe, K. Ahn,
  \emph{Small-scale Effects of Thermal Inflation on Halo Abundance at High-$z$, Galaxy Substructure Abundance and 21-cm Power Spectrum},
  Phys.\ Rev.\ D {\bf 96}, no. 10, 103515 (2017)
  [\href{https://arxiv.org/abs/1706.08049}{arXiv:1706.08049} [astro-ph.CO]].
\bibitem{2012MNRAS.421...50V} M. Viel, K. Markovic, M. Baldi, J. Weller, \emph{The Non-Linear Matter Power Spectrum in Warm Dark Matter Cosmologies}, Mon. Not. R. Astron. Soc. {\bf 421},  50-62 (2012)  [\href{https://arxiv.org/abs/1107.4094}{arXiv:1107.4094} [astro-ph.CO]].
\bibitem{Leo:2017wxg} M. Leo, C. M. Baugh, B. Li, S. Pascoli, \emph{Non-linear growth of structure in cosmologies with damped matter fluctuations}, (2017), [\href{https://arxiv.org/abs/1712.02742}{arXiv:1712.02742} [astro-ph.CO]].
\bibitem{2011arXiv1104.2932L} J. Lesgourgues, \emph{The Cosmic Linear Anisotropy Solving System (CLASS) I: Overview}, [\href{http://arxiv.org/abs/1104.2932}{arXiv:1104.2932} [astro-ph.IM]].
\bibitem{2011JCAP...09..032L} J. Lesgourgues, T. Tram, \emph{The Cosmic Linear Anisotropy Solving System (CLASS) IV: efficient implementation of non-cold relics}, JCAP {\bf 09}, 032 (2011) [\href{https://arxiv.org/abs/1104.2935}{arXiv:1104.2935}].
\bibitem{Lesgourgues:1997en} 
  J. Lesgourgues, D. Polarski, A. A. Starobinsky,
  \emph{CDM models with a BSI step - like primordial spectrum and a cosmological constant},
  Mon. Not. Roy. Astron. Soc.  {\bf 297}, 769 (1998) [\href{https://arxiv.org/abs/astro-ph/9711139}{arXiv:astro-ph/9711139}].  
\bibitem{Boyarsky:2008xj} A. Boyarsky, J. Lesgourgues, O. Ruchayskiy, M. Viel, \emph{Lyman-alpha constraints on warm and on warm-plus-cold
                        dark matter models}, JCAP {\bf 0905}, 012 (2009)   [\href{https://arxiv.org/abs/0812.0010v2}{arXiv:0812.0010} [astro-ph]].
\bibitem{Crocce:2006ve} M. Crocce, S. Pueblas, R. Scoccimarro, \emph{Transients from Initial Conditions in Cosmological Simulations}, Mon. Not. Roy. Astron. Soc.  {\bf 373}, 369 (2006)
[\href{https://arxiv.org/abs/astro-ph/0606505}{arXiv:astro-ph/0606505}].
\bibitem{Leo:2017zff} M. Leo, C. M. Baugh, B. Li, S. Pascoli, \emph{The Effect of Thermal Velocities on Structure Formation in N-body Simulations of Warm Dark Matter}, JCAP {\bf 11}, 017 (2017) [\href{https://arxiv.org/abs/1706.07837}{arXiv:1706.07837} [astro-ph.CO]].
\bibitem{Springel:2005mi} V. Springel, \emph{The cosmological simulation code GADGET-2}, Mon. Not. Roy. Astron. Soc.  {\bf 364}, 1105 (2005)
   [\href{http://arxiv.org/abs/astro-ph/0505010}{arXiv:astro-ph/0505010}].
\bibitem{Wang:2007he} J. Wang and S. D. M. White, \emph{Discreteness effects in simulations of Hot/Warm dark matter}, Mon. Not. Roy. Astron. Soc.  {\bf 380}, 93 (2007)
   [\href{https://arxiv.org/abs/astro-ph/0702575}{arXiv:astro-ph/0702575}].  
\bibitem{2012MNRAS.420.2318L} M. Lovell, V. Eke, C. Frenk, L. Gao, A. Jenkins, T. Theuns, J. Wang, S. White, A. Boyarsky, O. Ruchayskiy, \emph{The haloes of bright satellite galaxies in a warm dark matter universe}, Mon. Not. Roy. Astr. Soc. {\bf 420}, 2318-2324 (2012) [\href{https://arxiv.org/abs/1104.2929v2}{arXiv:1104.2929} [astro-ph.CO]].
\bibitem{2012MNRAS.424..684S} A. Schneider, R. E. Smith, A. V. Maccio, B. Moore, \emph{Non-linear evolution of cosmological structures in warm dark matter models}, Mon. Not. R. Astron. Soc. {\bf 424},  684-698 (2012)  [\href{https://arxiv.org/abs/1112.0330}{arXiv:1112.0330} [astro-ph.CO]].
\bibitem{Schneider:2013ria} 
  A. Schneider, R. E. Smith, D. Reed,
  \emph{Halo Mass Function and the Free Streaming Scale},
  Mon. Not. Roy. Astron. Soc.  {\bf 433}, 1573 (2013)
  [\href{https://arxiv.org/abs/1303.0839}{arXiv:1303.0839} [astro-ph.CO]].  
\bibitem{Lovell:2013ola} M. R. Lovell,  C. S. Frenk, V. R. Eke, A. Jenkins, L. Gao, T. Theuns, \emph{The properties of warm dark matter haloes}, Mon. Not. Roy. Astron. Soc.  {\bf 439}, 300 (2014)
   [\href{https://arxiv.org/abs/1308.1399}{arXiv:1308.1399} [astro-ph.CO]].  
\bibitem{Power:2013rpw} C. Power, 
  \emph{Seeking Observable Imprints of Small-Scale Structure on the Properties of Dark Matter Haloes},
   Publ. Astron. Soc. Austral.  {\bf 30}, 53 (2013)
   [\href{https://arxiv.org/abs/1309.1591v1}{arXiv:1309.1591} [astro-ph.CO]].
\bibitem{Schneider:2014rda} 
  A. Schneider,
  \emph{Structure formation with suppressed small-scale perturbations},
  Mon. Not. Roy. Astron. Soc.  {\bf 451}, no. 3, 3117 (2015)
  [\href{https://arxiv.org/abs/1412.2133}{arXiv:1412.2133} [astro-ph.CO]].  
\bibitem{Power:2016usj} C. Power, A. S. G. Robotham, D. Obreschkow, A. Hobbs, G. F. Lewis, 
  \emph{Spurious haloes and discreteness-driven relaxation in cosmological simulations},
   Mon. Not. Roy. Astron. Soc.  {\bf 462}, 474 (2016)
   [\href{https://arxiv.org/abs/1606.02038}{arXiv:1606.02038} [astro-ph.CO]].
\bibitem{Bose:2015mga} S. Bose, W. A. Hellwing, C. S. Frenk, A. Jenkins, M. R. Lovell, J. C. Helly, B. Li, \emph{The COpernicus COmplexio: Statistical Properties of Warm Dark Matter Haloes},  Mon. Not. Roy. Astron. Soc.  {\bf 455}, 318 (2016)
   [\href{https://arxiv.org/abs/1507.01998}{arXiv:1507.01998} [astro-ph.CO]].
\bibitem{2013ApJ...762..109B} P. S. Behroozi, R. H. Wechsler, H. Wu , \emph{The Rockstar Phase-Space Temporal Halo Finder and the Velocity Offsets of Cluster Cores}, ApJ {\bf 762}, 109 (2013)  [\href{https://arxiv.org/abs/1110.4372}{arXiv:1110.4372} [astro-ph.CO]].
\bibitem{Leo:2018odn} M. Leo, C. M. Baugh, B. Li, S. Pascoli, \emph{A new smooth-$k$ space filter approach to calculate halo abundances}, JCAP {\bf 04}, 010 (2018), [\href{https://arxiv.org/abs/1801.02547}{arXiv:1801.02547} [astro-ph.CO]].
\bibitem{Press:1973iz} 
  W. H. Press, P. Schechter,
  \emph{Formation of galaxies and clusters of galaxies by selfsimilar gravitational condensation},
  Astrophys. J.   {\bf 187}, 425 (1974).
\bibitem{Bond:1990iw} 
  J. R. Bond, S. Cole, G. Efstathiou, N. Kaiser,
  \emph{Excursion set mass functions for hierarchical Gaussian fluctuations},
  Astrophys. J.  {\bf 379}, 440 (1991).
\bibitem{Sheth:1999mn} 
  R. K. Sheth, G. Tormen,
  \emph{Large scale bias and the peak background split},
  Mon. Not.  Roy.  Astron.  Soc.   {\bf 308}, 119 (1999)
  [\href{https://arxiv.org/abs/astro-ph/9901122}{arXiv:astro-ph/9901122}].  
\bibitem{Zentner:2006vw} 
  A. R. Zentner,
  \emph{The Excursion Set Theory of Halo Mass Functions, Halo Clustering, and Halo Growth},
  Int. J. Mod. Phys. D {\bf 16}, 763 (2007) [\href{https://arxiv.org/abs/astro-ph/0611454v1}{arXiv:astro-ph/0611454}].
\bibitem{Maggiore:2009rv} 
  M. Maggiore, A. Riotto,
  \emph{The Halo Mass Function from Excursion Set Theory. I. Gaussian fluctuations with non-Markovian dependence on the smoothing scale},
  Astrophys. J.  {\bf 711}, 907 (2010)
  [\href{https://arxiv.org/abs/0903.1249}{arXiv:0903.1249} [astro-ph.CO]]. 
\bibitem{2013MNRAS.428.1774B} A. J. Benson, A. Farahi, S. Cole, L. A. Moustakas, A. Jenkins, M. Lovell, R. Kennedy, J. Helly, C. Frenk, \emph{Dark matter halo merger histories beyond cold dark matter - I. Methods and application to warm dark matter}, MNRAS {\bf 428}, 1774B (2013) [\href{https://arxiv.org/abs/1209.3018}{arXiv:1209.3018} [astro-ph.CO]].
\bibitem{Cyr-Racine:2015ihg} 
  F. Y. Cyr-Racine, K. Sigurdson, J. Zavala, T. Bringmann, M. Vogelsberger, C. Pfrommer,
  \emph{ETHOS - an effective theory of structure formation: From dark particle physics to the matter distribution of the Universe},
  Phys. Rev. D {\bf 93}, no. 12, 123527 (2016)
  [\href{https://arxiv.org/abs/1512.05344}{arXiv:1512.05344} [astro-ph.CO]].
\bibitem{Vogelsberger:2015gpr} 
  M. Vogelsberger, J. Zavala, F. Y. Cyr-Racine, C. Pfrommer, T. Bringmann, K. Sigurdson,
  \emph{ETHOS - an effective theory of structure formation: dark matter physics as a possible explanation of the small-scale CDM problems},
  Mon. Not. Roy. Astron. Soc. {\bf 460}, no. 2, 1399 (2016)
  [\href{https://arxiv.org/abs/1512.05349}{arXiv:1512.05349} [astro-ph.CO]].
\bibitem{Lovell:2015psz} 
  M. R. Lovell {\it et al.},
  \emph{Satellite galaxies in semi-analytic models of galaxy formation with sterile neutrino dark matter},
  Mon. Not. Roy. Astron. Soc.  {\bf 461}, no. 1, 60 (2016)
 [\href{https://arxiv.org/abs/1511.04078}{arXiv:1511.04078} [astro-ph.CO]].
\bibitem{Lovell:2017eec} 
  M. R. Lovell {\it et al.},
  \emph{ETHOS - an effective theory of structure formation: predictions for the high-redshift Universe - abundance of galaxies and reionization},
  Mon. Not. Roy. Astron. Soc.  {\bf 477}, no. 3, 2886 (2018)
  [\href{https://arxiv.org/abs/1711.10497}{arXiv:1711.10497} [astro-ph.CO]].
\bibitem{vanDaalen:2013ita} 
  M. P. van Daalen, J. Schaye, I. G. McCarthy, C. M. Booth, C. Dalla Vecchia,
  \emph{The impact of baryonic processes on the two-point correlation functions of galaxies, subhaloes and matter},
  Mon. Not. Roy. Astron. Soc.  {\bf 440}, no. 4, 2997 (2014)
  [\href{https://arxiv.org/abs/1310.7571}{arXiv:1310.7571} [astro-ph.CO]].
\bibitem{Vogelsberger:2014dza} 
  M. Vogelsberger {\it et al.},
  \emph{Introducing the Illustris Project: Simulating the coevolution of dark and visible matter in the Universe},
  Mon. Not. Roy. Astron. Soc.  {\bf 444}, no. 2, 1518 (2014)
  [\href{https://arxiv.org/abs/1405.2921}{arXiv:1405.2921} [astro-ph.CO]].    
\bibitem{Sawala:2015cdf} 
  T. Sawala {\it et al.},
  \emph{The APOSTLE simulations: solutions to the Local Group's cosmic puzzles},
  Mon. Not. Roy. Astron. Soc.  {\bf 457}, no. 2, 1931 (2016)
  [\href{https://arxiv.org/abs/1511.01098}{arXiv:1511.01098} [astro-ph.GA]].
\bibitem{Hellwing:2016ucy} 
  W. A. Hellwing, M. Schaller, C. S. Frenk, T. Theuns, J. Schaye, R. G. Bower, R. A. Crain,
  \emph{The effect of baryons on redshift space distortions and cosmic density and velocity fields in the EAGLE simulation},
  Mon. Not. Roy. Astron. Soc.  {\bf 461}, no. 1, L11 (2016)
  [\href{https://arxiv.org/abs/1603.03328v2}{arXiv:1603.03328} [astro-ph.CO]].
 \bibitem{Artale:2016rit} 
  M. C. Artale {\it et al.},
  \emph{Small-scale galaxy clustering in the EAGLE simulation},
  Mon. Not. Roy. Astron. Soc.  {\bf 470}, no. 2, 1771 (2017)
  [\href{https://arxiv.org/abs/1611.05064}{arXiv:1611.05064} [astro-ph.GA]].
\bibitem{2017MNRAS.468.4285L} M. R. Lovell {\it et al.}, \emph{Properties of Local Group galaxies in hydrodynamical simulations of sterile neutrino dark matter cosmologies}, Mon. Not. Roy. Astron. Soc. {\bf 468}, 4285-4298 (2017) [\href{https://arxiv.org/abs/1611.00010}{arXiv:1611.00010} [astro-ph.GA]].
\bibitem{Springel:2017tpz} 
  V. Springel {\it et al.},
  \emph{First results from the IllustrisTNG simulations: matter and galaxy clustering},
  Mon. Not. Roy. Astron. Soc.  {\bf 475}, 676 (2018)
   [\href{https://arxiv.org/abs/1707.03397}{arXiv:1707.03397} [astro-ph.GA]].
\bibitem{2018Natur.555...67B} J. D. Bowman, A. E. E. Rogers, R. A. Monsalve, T. J. Mozdzen, N. Mahesh, \emph{An absorption profile centred at 78 megahertz in the sky-averaged spectrum}, Nature {\bf 555}, 67–70 (2018). 
\bibitem{Schneider:2018xba} 
  A. Schneider,
  \emph{Constraining Non-Cold Dark Matter Models with the Global 21-cm Signal}, (2018)
  [\href{https://arxiv.org/abs/1805.00021}{arXiv:1805.00021} [astro-ph.CO]].
\bibitem{Lidz:2018fqo} 
  A. Lidz, L. Hui,
  \emph{The Implications of a Pre-reionization 21 cm Absorption Signal for Fuzzy Dark Matter}, (2018)
  [\href{https://arxiv.org/abs/1805.01253v1}{arXiv:1805.01253} [astro-ph.CO]].
\bibitem{Viel:2013apy} M. Viel, G. D. Becker, J. S. Bolton, M. G. Haehnelt, \emph{Warm dark matter as a solution to the small scale
                        crisis: New constraints from high redshift Lyman-$\alpha$ forest
                        data}, Phys. Rev. D {\bf 88}, 043502 (2013)  [\href{http://arxiv.org/abs/1306.2314v2}{arXiv:1306.2314} [astro-ph.CO]].
\bibitem{Irsic:2017ixq} V. Irsic {\it et al.}, \emph{New Constraints on the free-streaming of warm dark matter from intermediate and small scale Lyman-$\alpha$ forest data}, [\href{https://arxiv.org/abs/1702.01764}{arXiv:1702.01764} [astro-ph.CO]].
\bibitem{Gardner:2006ky} 
  J. P. Gardner {\it et al.},
  \emph{The James Webb Space Telescope},
  Space Sci. Rev.  {\bf 123}, 485 (2006)
[\href{https://arxiv.org/abs/astro-ph/0606175}{arXiv:astro-ph/0606175}].
\bibitem{2016SPIE.9908E..1ZD} 
  R. Davies {\it et al.},
  \emph{MICADO: first light imager for the E-ELT},
  Ground-based and Airborne Instrumentation for Astronomy VI {\bf 9908}, 99081Z (2016)
[\href{https://arxiv.org/abs/1607.01954}{arXiv:1607.01954} [astro-ph.IM]].
\bibitem{Li:2015xpc} 
  R. Li, C. S. Frenk, S. Cole, L. Gao, S. Bose, W. A. Hellwing,
  \emph{Constraints on the identity of the dark matter from strong gravitational lenses},
  Mon. Not. Roy. Astron. Soc. {\bf 460}, 363 (2016)
  [\href{https://arxiv.org/abs/1512.06507}{arXiv:1512.06507} [astro-ph.CO]].
  



\end{thebibliography}
\end{document}